\documentclass[twoside,twocolumn,12pt]{article}
\usepackage{epsfig}

\newcommand{\be}{\begin{equation}}
\newcommand{\ee}{\end{equation}}
\newcommand{\bea}{\begin{eqnarray}}
\newcommand{\eea}{\end{eqnarray}}

\begin{document}

\title{ \vspace{1cm} Studying the Earth with Geoneutrinos}

\title{ \vspace{1cm} Studying the Earth with Geoneutrinos}

\author{L.\ Ludhova,$^1$ S.\ Zavatarelli$^2$\\
\\
$^1$Dipartimento di Fisica, INFN, 20133 Milano, Italy\\
$^2$Dipartimento di Fisica, INFN, 16146 Genova, Italy
}

\maketitle
\begin{abstract} 
Geo-neutrinos, electron antineutrinos from natural radioactive decays inside the Earth, bring to the surface unique information about our planet. The new techniques in neutrino detection opened a door into a completely new inter-disciplinary field of Neutrino Geoscience. We give here a broad geological introduction highlighting the points where the geo-neutrino measurements can give substantial new insights. The status-of-art of this field is overviewed, including a description of the latest experimental results from KamLAND and Borexino experiments and their first geological implications. We performed a new combined Borexino and KamLAND analysis in terms of the extraction of the mantle geo-neutrino signal and the limits on the Earth's radiogenic heat power. The perspectives and the future projects having geo-neutrinos among their scientific goals are also discussed. 
\end{abstract}
\

\section{Introduction}
\label{Sec:Intro}

The newly born inter-disciplinar field of Neutrino Geoscience takes the advantage of the technologies developed by large-volume neutrino experiments and of the achievements of the elementary particle physics in order to study the Earth interior with a new probe - geo-neutrinos. Geo-neutrinos are electron antineutrinos released in the decays of radioactive elements with lifetimes comparable with the age of the Earth and distributed through the Earth's interior. The radiogenic heat released during the decays of these Heat Producing Elements (HPE) is in a well fixed ratio with the total mass of HPE inside the Earth. Geo-neutrinos bring to the Earth's surface an instant information about the distribution of HPE. Thus, it is, in principle, possible to extract from measured geo-neutrino fluxes several geological information completely unreachable by other means. These information concern the total abundance and distribution of the HPE inside the Earth and thus the determination of the fraction of radiogenic heat contributing to the total surface heat flux. Such a knowledge is of critical importance for understanding complex processes such as the mantle convection, the plate tectonics, the geo-dynamo (the process of generation of the Earth's magnetic field), as well as the Earth  formation itself. 

Currently, only two large-volume, liquid-scintillator neutrino experiments, KamLAND in Japan and Borexino in Italy, have been able to measure the geo-neutrino signal. Antineutrinos can interact only through the weak interactions. Thus,  the cross-section of the inverse-beta decay detection interaction
\begin{equation}
\bar{\nu}_e + p \rightarrow e^+ + n
\label{Eq:InvBeta}
\end{equation}
is very low. Even a typical flux of the order of $10^{6}$ geo-neutrinos cm$^{-2}$ s$^{-1}$ leads to only a hand-full number of interactions, few or few tens per year with the current-size detectors. This means, the geo-neutrino experiments must be installed in underground laboratories in order to shield the detector from cosmic radiations.

The aim of the present paper is to review the current status of the Neutrino Geoscience. First, in Sec.~\ref{Sec:Geonu} we describe the radioactive decays of HPE and the geo-neutrino production, the geo-neutrino energy spectra and the impact of the neutrino oscillation phenomenon on the geo-neutrino spectrum and flux. The Sec.~\ref{Sec:Earth} is intended to give an overview of the current knowledge of the Earth interior. The opened problems to which understanding the geo-neutrino studies can contribute are highlighted. Section~\ref{Sec:signal} sheds light on how the expected geo-neutrino signal can be calculated considering different geological models. Section~\ref{Sec:experiments} describes the KamLAND and the Borexino detectors. Section~\ref{Sec:analysis} describes details of the geo-neutrino analysis: from the detection principles, through the background sources to the most recent experimental results and their geological implications. Finally, in Sec.~\ref{Sec:future} we describe the future perspectives of the field of Neutrino Geoscience and the projects having geo-neutrino measurement among their scientific goals.

\section{Geo-neutrinos}
\label{Sec:Geonu}

Today, the Earth's radiogenic heat is in almost 99\%  produced along with the radioactive decays in the chains of  $^{232}$Th ($\tau _{1/2}$ = 14.0 $\cdot $ 10$^{ 9}$ year), $^{238}$U ($\tau _{1/2}$ = 4.47 $\cdot $ 10$^{ 9}$ year), $^{235}$U ($\tau _{1/2}$ = 0.70 $\cdot $ 10$^{ 9}$ year),  and those of the $^{40}$K isotope ($\tau _{1/2}$ = 1.28 $\cdot $ 10$^{ 9}$ year).
The overall decay schemes and the heat released in each of these decays are summarized in the following equations:

\begin{equation}
^{238}\mathrm{U} \rightarrow ^{206}\mathrm{Pb} + 8\alpha + 8 e^{-} + 6 \bar{\nu}_e + 51.7 ~~~\mathrm{MeV} 
\label{Eq:geo1}
\end{equation}
\begin{equation}
^{235}\mathrm{U} \rightarrow ^{207}\mathrm{Pb} + 7\alpha + 4 e^{-} + 4 \bar{\nu}_e + 46.4 ~\mathrm{MeV} 
\label{Eq:geo2}
\end{equation}
 \begin{equation}
^{232}\mathrm{Th} \rightarrow ^{208}\mathrm{Pb} + 6\alpha + 4 e^{-} + 4 \bar{\nu}_e + 42.7 ~\mathrm{MeV} 
\label{Eq:geo3}
\end{equation}
 \begin{equation}
^{40}\mathrm{K} \rightarrow ^{40}\mathrm{Ca}  +  e^{-} +  \bar{\nu}_e + 1.31 ~\mathrm{MeV}~\mathrm{(89.3\%)}
\label{Eq:geo4}
\end{equation}
 \begin{equation}
^{40}\mathrm{K} + \mathrm{e}  \rightarrow ^{40}\mathrm{Ar}  +  \nu _{e} +  1.505 ~\mathrm{MeV}~~~\mathrm{(10.7\%)}
\label{Eq:geo5}
\end{equation}
Since the isotopic abundance of $^{235}$U is small, the overall contribution of $^{238}$U, $^{232}$Th, and $^{40}$K is largely predominant.
In addition, a small fraction (less than 1\%) of the radiogenic heat is coming from the decays of $^{87}$Rb ($\tau _{1/2}$ = 48.1 $\cdot$ 10$^{ 9}$ year), $^{138}$La ($\tau _{1/2}= 102 \cdot$ 10$^{ 9}$ year), and $^{176}$Lu ($\tau _{1/2}$ = 37.6 $\cdot$ 10$^{ 9}$ year).

Neutron-rich nuclides like $^{238}$U, $^{232}$Th, and $^{235}$U, made up~\cite{rolfs} by neutron capture reactions during the last stages of massive-stars lives, decay into the lighter and proton-richer nuclides by yielding $\beta^{-}$ and $\alpha$ particles, see Eqs.~\ref{Eq:geo1} - \ref{Eq:geo3}. During $\beta ^{-}$ decays,  electron antineutrinos ($\bar{\nu}_e$) are emitted that carry away in the case of $^{238}$U and $^{232}$Th chains,  8\% and 6\%, respectively, of the total available energy ~\cite{fiorentini2007}.
In the computation of the overall $\bar{\nu}_e$ energy spectrum of each decay chain, the shapes and rates of all the individual decays has to be included: detailed calculations required to take into account up to $\sim$80 different branches for each chain~\cite{enomotothesis}.
The most important contributions to the geo-neutrino signal are however those of $^{214}$Bi and $^{234}$Pa$\rm^{m}$ in the uranium chain and $^{212}$Bi and $^{228}$Ac in the thorium chain~\cite{fiorentini2007}.

\begin{figure}[tb]
\begin{center}
\centering{\epsfig{file=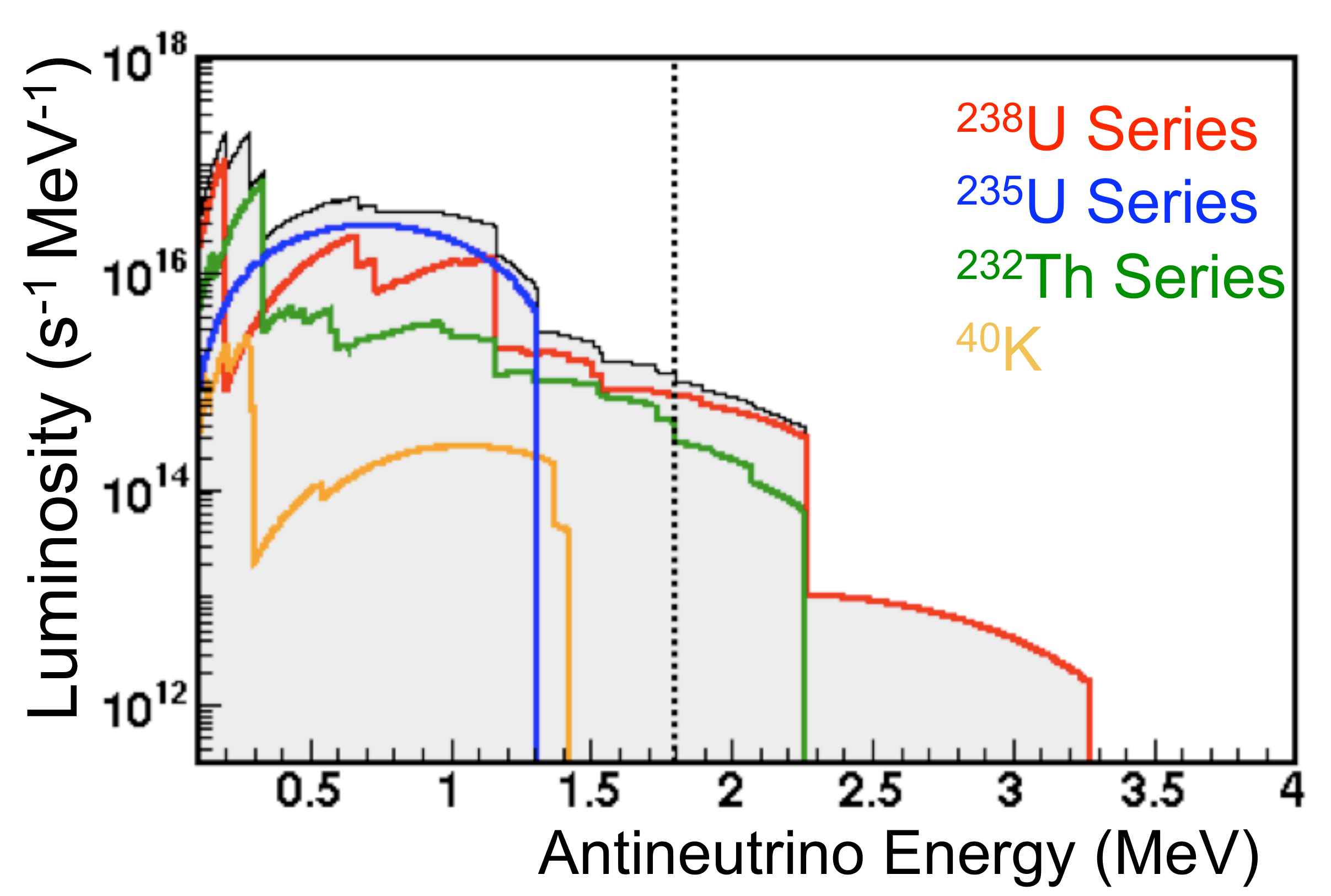,scale=0.38}}
\caption{The geo-neutrino luminosity as a function of energy is shown for the most important reaction chains and nuclides~\cite{enomototalk}. Only geo-neutrinos of energies above the 1.8\,MeV energy (vertical dashed line) can be detected by means of the inverse beta decay on target protons shown in Eq.~\ref{Eq:InvBeta}.}
\label{Fig:geonuspe}
\end{center}
\end{figure}

Geo-neutrino spectrum extends up to 3.26\,MeV and the contributions originating from different elements can be distinguished according to their different end-points, i.e., geo-neutrinos with E $\textgreater$ 2.25\,MeV are produced only in the uranium chain, as shown in Fig.~\ref{Fig:geonuspe}. We note, that according to geo-chemical studies, $^{232}$Th is more abundant than $^{238}$U and their mass ratio in the bulk Earth is expected to be $m$($^{232}$Th)/$m$($^{238}$U) = 3.9 (see also Sec.~\ref{Sec:Earth}). Because the cross-section of the detection interaction from Eq.~\ref{Eq:InvBeta} increases with energy, the ratio of the signals expected in the detector is $S$($^{232}$Th)/$S$($^{238}$U) = 0.27.  

The $^{40}$K nuclides presently contained in the Earth were formed during an earlier and more quiet phase of the massive-stars evolution, the so called Silicon burning phase~\cite{rolfs}. In this phase, at temperatures higher than 3.5 $\cdot$ 10$^{ 9}$ K, $\alpha$ particles, protons, and neutrons were ejected by photo-disintegration from the nuclei abundant in these stars and were made available for building-up the light nuclei up to and slightly beyond the iron peak (A = 65).
Being a lighter nucleus, the $^{40}$K, beyond the $\beta ^{-}$ decay shown in Eq.~\ref{Eq:geo4}, has also a sizeable decay branch (10.7\%) by electron capture, see Eq.~\ref{Eq:geo5}. In this case, electron neutrinos are emitted but they are not easily observable because they are overwhelmed by the many orders of magnitude more intense solar-neutrino fluxes.
Luckily, the Earth is mostly shining in antineutrinos; the Sun, conversely, is producing energy by light-nuclide fusion reactions  and only neutrinos are yielded during such processes.

Both the $^{40}$K and $^{235}$U geo-neutrinos are below the 1.8\,MeV threshold  of Eq.~\ref{Eq:InvBeta}, as shown in Fig.~\ref{Fig:geonuspe}, and thus, they cannot be detected by this process. 
However, the elemental abundances ratios are much better known than the absolute abundances. Therefore, by measuring the absolute content of $^{238}$U and $^{232}$Th, also the overall amount of $^{40}$K and $^{235}$U can be inferred with an improved precision.

Geo-neutrinos are emitted and interact as flavor states but they do travel as superposition of mass states and are therefore subject to flavor oscillations.

In the approximation $\Delta m^{2}_{31}$ $\sim$ $\Delta m^{2}_{32}$ $\gg$ $\Delta m^{2}_{21}$, the square-mass differences of mass eigenstates 1, 2, and 3, the survival probability $P_{ee}$ for a $\bar{\nu}_e$ in vacuum is:

 \begin{eqnarray}
 P_{ee} = P ( \bar{\nu}_e \rightarrow \bar{\nu}_e) = \sin ^{4} \theta_{13}+ \nonumber \\
  +  \cos ^{4} \theta_{13}  \left(1-  \sin^{2}2  \theta _{12}  sin^{2}\left(\frac { 1.267~\Delta m^{2}_{21} L}{4E}\right) \right)  
 \label{Eq:geo6}
\end{eqnarray}

In the Earth, the geo-neutrino sources are spread over a vast region compared to the oscillation length: 
 \begin{equation}
L  \sim \pi c \hbar  \frac{4E} {\Delta m^{2}_{21}} 
\label{Eq:geo7}
\end{equation}
For example, for a $\sim$3\,MeV antineutrino, the oscillation length is of $\sim$100\,km, small with respect to the Earth's radius of $\sim$6371\,km: the effect of the neutrino oscillation to the total neutrino flux is well averaged, giving an overall survival probability of:
\begin{equation}
\langle {P_{ee}} \rangle  \simeq \cos ^{4} \theta_{13} \left(1- \frac{1} {2}  \sin ^{2} 2 \theta _{12}\right)   + \sin ^{4} \theta _{13}
\label{Eq:geo8}
\end{equation}
According to the neutrino oscillation mixing angles and square-mass differences reported in~\cite{fogli2012}, $P_{ee}\sim 0.54$. 

While geo-neutrinos propagate through the Earth, they feel the potential of electrons and nucleons building-up the surrounding matter. The charged weak current interactions affect only the electron flavor (anti)neutrinos. As a consequence, the Hamiltonian for $\bar{\nu}_e$'s has an extra term of $\sqrt{2} G_{F} n_{e}$, where $n_{e}$ is the electron density. Since the electron density in the Earth is not constant and moreover it shows sharp changes in correspondence with boundaries of different Earth's layers, the behavior of the survival probability is not trivial and the motion equations have to be solved by numerical tracing. It has been calculated in~\cite{enomotothesis} that this so called {\it matter effect} contribution to the average survival probability is an increase of about 2\% and the spectral distortion is below 1\%.

To conclude, the net effect of flavor oscillations during the geo-neutrino ($\bar{\nu}_e$) propagation through the Earth is the absolute decrease of the overall flux by $\sim$0.55 with a very small spectral distortion, negligible for the precision of the current geo-neutrino experiments.

\section{The Earth}
\label{Sec:Earth}

The Earth was created in the process of accretion from undifferentiated material, to which chondritic meteorites are believed to be the closest in composition and structure.
The Ca-Al rich inclusions in carbonaceous chondrite meteorites up to about a cm in size are the oldest known solid condensates from the hot material of the protoplanetary disk.
The age of these fine grained structures was determined based on U-corrected Pb-Pb dating to be $4567.30 \pm 0.16$ million years~\cite{connely}. Thus, these inclusions together with the so called chondrules, another type of inclusions of similar age, provide an upper limit on the age of the Earth. The oldest terrestrial material are zircon inclusions from Western Australia being at least 4.404 billion years old~\cite{wild}.

The bodies with a sufficient mass undergo the process of differentiation, e. g., a transformation from an homogeneous object to a body with a layered structure. 
The metallic core of the Earth (and presumably also of other terrestrial planets) was the first to differentiate during the first $\sim$30~million years of the life of the Solar System, as inferred based on the $^{182}$Hf - $^{182}$W isotope system~\cite{kleine}. Today, the core has a radius of 2890\,km, about 45\% of the Earth radius and represents less than 10\% of the total Earth volume. 

Due to the high pressure of about 330\,GPa, the Inner Core with 1220\,km radius is solid, despite the high temperature of $\sim$5700\,K, comparable to the temperature of the solar photosphere. 

From seismologic studies, and namely from the fact that the secondary, transverse/shear waves do not propagate through the so called Outer Core, we know that it is liquid. Turbulent convection occurs in this liquid metal of low viscosity. These movements have a crucial role in the process of the generation of the Earth magnetic field, so called geo-dynamo. The magnetic field here is about 25\,Gauss, about 50 times stronger than at the Earth's surface. 

The chemical composition of the core is inferred indirectly as Fe-Ni alloy with up to 10\% admixture of light elements, most probable being oxygen and/or sulfur. Some high-pressure, high-temperature experiments confirm that potassium enters iron sulfide melts in a strongly temperature-dependent fashion and that $^{40}$K could thus serve as a substantial heat source in the core~\cite{murthy}. However, other authors show that several geo-chemical arguments are not in favor of such hypothesis~\cite{mcdonough}.  Geo-neutrinos from $^{40}$K have energies below the detection threshold of the current detection method (see Fig.~\ref{Fig:geonuspe}) and thus the presence of potassium in the core cannot be tested with geo-neutrino studies based on inverse beta on free protons. Other heat producing elements, such as uranium and thorium are lithophile elements and due to their chemical affinity they are quite widely believed not to be present in the core (in spite of their high density). There exist, however, ideas as that of Herndon~\cite{herndon} suggesting an U-driven georeactor with thermal power $<$30\,TW present in the Earth's core and confined in its central part within the radius of about 4\,km.  The antineutrinos that would be emitted from such a hypothetical georeactor have, as antineutrinos from the nuclear power plants, energies above the end-point of geo-neutrinos from "standard" natural radioactive decays. Antineutrino detection provide thus a sensitive tool to test the georeactor hypothesis.

After the separation of the metallic core, the rest of the Earth's volume was composed by a presumably homogeneous Primitive Mantle built of silicate rocks which subsequently differentiated to the present mantle and crust. 

Above the Core Mantle Boundary (CMB) there is a $\sim$200\,km thick zone called D'' (pronounced D-double prime), a seismic discontinuity characterized by a decrease in the gradient of both P (primary) and S (secondary, shear) wave velocities. The origin and character of this narrow zone is under discussion and there is no widely accepted model.

The Lower Mantle is about 2000\,km thick and extends from the D'' zone up to the seismic discontinuity at the depth of 660\,km. This discontinuity does not represent a chemical boundary while a zone of a phase transition and mineral recrystallization. Below this zone, in the Lower Mantle, the dominant mineral phases are the Mg-perovskite (Mg$_{0.9}$Fe$_{0.1}$)SiO$_3$, ferropericlase (Mg,Fe)O, and Ca-perovskite CaSiO$_3$. The temperature at the base of the mantle can reach 3700\,K while at the upper boundary the temperature is about 600\,K. In spite of such high temperatures, the high lithostatic pressure (136\,GPa at the base) prevents the melting, since the solidus increases with pressure. The Lower Mantle is thus solid, but viscose and undergoes plastic deformation on long time-scales. Due to a high temperature gradient and the ability of the mantle to creep, there is an ongoing convection in the mantle. This convection drives the movement of tectonic plates with characteristic velocities of few cm per year. The convection may be influenced by the mineral recrystallizations occurring at 660\,km and 410\,km depths, through the density changes and latent heat. 

The mantle between these two seismic discontinuities at 410 and 660\,km depths is called the Transition Zone. This zone consists primarily of peridotite rock with dominant minerals garnet (mostly pyrop Mg$_3$Al$_2$(SiO$_4$)$_3$) and high-pressure polymorphs of olivine (Mg, Fe)$_2$SiO$_4$, ringwoodite and wadsleyite below and above cca. 525\,km depth, respectively. 

In the Upper Mantle above the 410\,km depth discontinuity the dominant minerals are olivine, garnet, and pyroxene. The upper mantle boundary is defined with seismic discontinuity called Mohorovi\v{c}i\'c, often referred to as Moho. It's average depth is about 35\,km, 5 - 10\,km below the oceans and 20 - 90\,km below the continents. The Moho lies within the lithosphere, the mechanically defined uppermost Earth layer with brittle deformations composed of the crust and the brittle part of the upper mantle, Continental Lithospheric Mantle (CLM). The lithospheric tectonic plates are floating on the more plastic astenosphere entirely composed of the mantle material. 

Partial melting is a process when solidus and liquidus temperatures are different and are typical for heterogeneous systems as rocks. The mantle partial melting through geological times lead to the formation of the Earth's crust. Typical mantle rocks have a higher magnesium-to-iron ratio and a smaller proportion of silicon and aluminum than the crust. The crust can be seen as the accumulation of solidified partial liquid, which thanks to its lower density tends to move upwards with respect to denser solid residual. The lithophile and incompatible elements, such as U and Th, tend to concentrate in the liquid phase and thus they do concentrate in the crust. 

There are two types of the Earth's crust. The simplest and youngest is the oceanic crust, less than 10\,km thick. It is created by partial melting of the Transition-Zone mantle along the mid-oceanic ridges on top of the upwelling mantle plumes. The total length of this submarine mountain range, the so called rift zone, is about 80,000\,km. The age of the oceanic crust is increasing with the perpendicular distance from the rift, symmetrically on both sides. The oldest large-scale oceanic crust is in the west Pacific and north-west Atlantic - both are up to 180 - 200 million years old.   However, parts of the eastern Mediterranean Sea are remnants of the much older Tethys ocean, at about 270 million years old. The typical rock types of the oceanic crust created along the rifts are Mid-Ocean Ridge Basalts (MORB). They are relatively enriched in lithophile elements with respect to the mantle from which they have been differentiated but they are much depleted in them with respect to the continental crust. The typical density of the oceanic crust is about 2.9\,g cm$^{-3}$.

The continental crust is thicker, more heterogeneous and older, and has a more complex history with respect to the oceanic crust. It forms continents and continental shelves covered with shallow seas. The  bulk composition is granitic, more felsic with respect to oceanic crust. Continental crust covers about 40\% of the Earth surface. It is much thicker than the oceanic crust, from 20 to 70\,km. The average density is 2.7\,g cm$^{-3}$, less dense than the oceanic crust and so to the contrary of the oceanic crust, the continental slabs rarely subduct. Therefore, while the subducting oceanic crust gets destroyed and remelted, the continental crust persists. On average, it has about 2 billion years, while the oldest rock is the Acasta Gneiss from the continental root (craton) in Canada is about 4 billion years old. The continental crust is thickest in the areas of continental collision and compressional forces, where new mountain ranges are created in the process called orogeny, as in the Himalayas or in the Alps. There are the three main rock groups building up the continental crust: igneous (rocks which solidified from a molten magma (below the surface) or lava (on the surface)), sedimentary (rocks that were created by the deposition of  the material as disintegrated older rocks, organic elements etc.), and metamorphic (rocks that recrystallized without melting under the increased temperature and/or pressure conditions).

There are several ways in which we can obtain information about the deep Earth. Seismology studies the propagation of the P (primary, longitudinal) and the S (secondary, shear, transversal) waves through the Earth and can construct the wave velocities and density profiles of the Earth. It can identify the discontinuities corresponding to mechanical and/or compositional boundaries. The first order structure of the Earth's interior is defined by the 1D seismological profile, called PREM: Preliminary Reference Earth Model~\cite{dziewonski}. The recent seismic tomography can reveal structures as 
Large Low Shear Velocity Provinces (LLSVP) below Africa and central Pacific~\cite{wang} indicating that mantle could be even compositionally non-homogeneous and that it could be tested via future geo-neutrino projects~\cite{sramek}.

The chemical composition of the Earth is the subject of study of geochemistry. The direct rock samples are however limited. The deepest bore-hole ever made is 12\,km in Kola peninsula in Russia. Some volcanic and tectonic processes can bring to the surface samples of deeper origin but often their composition can be altered during the transport. The pure samples of the lower mantle are practically in-existent. With respect to the mantle, the composition of the crust is relatively well known. A comprehensive review of the bulk compositions of the upper, middle, and lower crust were published by Rudnick and Gao~\cite{rudnick} and Huang et al.~\cite{huang}. 

The bulk composition of the silicate Earth, the so called Bulk Silicate Earth (BSE) models describe the composition of the Primitive Mantle, the Earth composition after the core separation and before the crust-mantle differentiation. The estimates of the composition of the present-day mantle can be derived as a difference between the mass abundances predicted by the BSE models in the Primitive Mantle and those observed in the present crust. In this way, the predictions of the U and Th mass abundances in the mantle are made, which are then critical in calculating the predicted geo-neutrino signal, see Sec.~\ref{Sec:signal}.

The refractory elements are those that have high condensation temperatures; thus, they did condensate from a hot nebula, today form the bulk mass of the terrestrial planets, and are observed in equal proportions in the chondrites. Their contrary are volatile elements with low condensation temperatures and which might have partially escaped from the planet. U and Th are refractory elements, while K is moderately volatile. All U, Th, and K are also lithophile (rock-loving) elements, which in the  Goldschmidt geochemical classification means elements tending to stay in the silicate phase (other categories are siderophile (metal-loving), chalcophile (ore, chalcogen-loving), and atmophile/volatile). 

The most recent classification of BSE models was presented by \v{S}r\'amek et al.~\cite{sramek}:
\begin{itemize}
\item{{\it Geochemical BSE models:} these models rely on the fact that the composition of carbonaceous (CI) chondrites matches the solar photospheric abundances in refractory lithophile, siderophile, and volatile elements. These models assume that the ratios of Refractory Lithophile Elements (RLE) in the bulk silicate Earth are the same as in the CI chondrites and in the solar photosphere. The typical chondritic value of the bulk mass Th/U ratio is 3.9 and K/U $\sim$ 13,000. The absolute RLE abundances are inferred from the available crust and upper mantle rock samples. The theoretical petrological models and melting trends are taken into account in inferring the composition of the original material of the Primitive Mantle, from which the current rocks were derived in the process of partial melting. Among these models are McDonough and Sun (1995)~\cite{McDonoughSun}, All\'egre (1995)~\cite{allegre95}, Hart and Zindler (1986)~\cite{hart}, Arevalo et al. (2009)~\cite{arevalo}, and Palme and O'Neill (2003)~\cite{palme}. The typical U concentration in the bulk silicate Earth is about $20 \pm 4$\,ppb.}
\item{{\it Cosmochemical BSE models:}  The model of Javoy et al. (2010)~\cite{javoy} builds the Earth from the enstatite chondrites, which show the closest isotopic similarity with mantle rocks and have sufficiently high iron content to explain the metallic core (similarity in oxidation state). The "collisional erosion" model of O'Neill and Palme (2008)~\cite{ONeill} is covered in this category as well. In this model, the early enriched crust was lost in the collision of the Earth with an external body. In both of these models the typical bulk U concentration is about 10-12 ppb. }
\item{{\it Geodynamical BSE models:} These models are based on the energetics of the mantle convection. Considering the current surface heat flux, which depends on the radiogenic heat and the secular cooling, the parametrized convection models require higher contribution of radiogenic heat (and thus higher U and Th abundances) with respect to geo and cosmochemical models. The typical bulk U concentration is $35 \pm 4$\,ppb.}
\end{itemize}   

The surface heat flux is estimated based on the measurements of temperature gradients along several thousands of drill holes along the globe. The   most recent evaluation of these data leads to the prediction of $47 \pm 2$\,TW predicted by Davies and Davies (2010)~\cite{davies}, consistent with the estimation of Jaupart et al. (2007)~\cite{jaupart}. The relative contribution of the radiogenic heat from radioactive decays to this flux (so called Urey ratio) is not known and this is the key information which can be pinned down by the geo-neutrino measurements. The geochemical, cosmochemical, and geodynamical models predict the radiogenic heat of 20 $\pm$ 4, 11 $\pm$ 2, 33 $\pm$ 3\,TW and the  corresponding Urey ratios of about 0.3, 0.1, and 0.6, respectively. The Heat Producing Elements (HPE) predicted by these models are distributed in the crust and in the mantle. The crustal radiogenic power was recently evaluated by Huang et al.~\cite{huang} as $6.8^{+1.4}_{-1.1}$\,TW. By subtracting this contribution from the total radiogenic heat predicted by different BSE models, the mantle radiogenic power driving the convection and plate tectonics can be as little  as 3\,TW and as much as 23\,TW. To determine this mantle contribution is one of the main goals and potentials of Neutrino Geoscience.

\section{Geo-neutrino signal prediction}
\label{Sec:signal}

The geo-neutrino signal can be expressed in several ways. We recall that geo-neutrinos are detected by the inverse beta decay reaction (see Eq.~\ref{Eq:InvBeta}) in which antineutrino interacts with a target proton. The most straightforward unit is the normalized event rate, expressed by the so called Terrestrial Neutrino Unit (TNU), defined as the number of interactions detected during one year on a target of  $10^{32}$ protons ($\sim$1\,kton of liquid scintillator) and with 100\% detection efficiency. Conversion between the signal $S$ expressed in TNU and the oscillated, electron flavor flux $\phi$ (expressed in $10^6  \rm{cm}^{-2} \rm{s}^{-1}$) is straightforward~\cite{lisi} and requires a knowledge of the geo-neutrino energy spectrum and the interaction cross section, which scales with the $\bar{\nu}_e$ energy:
 \begin{equation}
S(^{232}\rm{Th}) [\rm{TNU}] = 4.07 \cdot \phi (^{232}\rm{Th})
\label{Eq:TNUFluxTh}
\end{equation}
 \begin{equation}
S(^{238}\rm{U}) [\rm{TNU}] = 12.8 \cdot \phi (^{238}\rm{U})
\label{Eq:TNUFluxU}
\end{equation}

In order to calculate the geo-neutrino signal at a certain location on the Earth's surface, it is important to know the absolute amount and the distribution of HPE inside the Earth. As it was described in Sec.~\ref{Sec:Earth}, we know relatively well such information for the Earth's crust, but we lack it for the mantle. Instead, the BSE models, also described in Sec.~\ref{Sec:Earth}, predict the total amount of HPE in the silicate Earth (so, excluding the metallic core, in which no HPE are expected). Thus, in the geo-neutrino signal predictions, the procedure is as follows. First, the signal from the crust is calculated. Then, the total mass of the HPE concentrated in the crust is subtracted from the HPE mass predicted by a specific BSE model; the remaining amount of HPE is attributed to be concentrated in the mantle.

Due to the chemical affinity of HPE, the continental crust is their richest reservoir. Thus, for the experimental sites built on the continental crust, the total geo-neutrino signal is dominated by the crustal component. It is important to estimate it with the highest possible precision since the mantle contribution can be extracted from the measured signal only after the subtraction of the expected crustal component.  

The first estimation of the crustal geo-neutrino signal~\cite{krauss} modeled the crust as a homogeneous, 30\,km thick layer. Since then, several much more refined models have been developed. In these models, different geochemical and geophysical data are used as input parameters. The crust is divided in finite volume voxels with surface area of either $5\rm{^{\circ}} \times 5\rm{^{\circ}}$~\cite{calaprice},  $2\rm{^{\circ}} \times 2\rm{^{\circ}}$~\cite{enomoto, fogli2006, mantovani2004} or, most recently, $1\rm{^{\circ}} \times 1\rm{^{\circ}}$~\cite{huang}. The oceanic and continental crust are treated separately. The continental crust is further divided in different layers, as upper, middle, and lower continental crust. 

On the sites placed on the continental crust, a significant part of the crustal signal comes from the area with a radius of few hundreds of km around the detector~\cite{mantovani2004}. Thus, in a precise estimation of the crustal geo-neutrino signal, it is important to distinguish the contribution from the local crust (LOC) and the rest of the crust (ROC)~\cite{fiorentini2012}. In estimating the LOC contribution, it is crucial to consider the composition of real rocks surrounding the experimental site, while for the ROC contribution it is sufficient to take into account the mean crustal compositions.

Borexino and KamLAND, the only two experiments which have provided geo-neutrino measurements, are placed in very different geological environments. Borexino is placed on a continental crust in central Italy. KamLAND is situated in Japan, in an area with very complicated geological structure around the subduction zone. In Table~\ref{tab:signalExp} we show the expected geo-neutrino signal for both experiments. 

\begin{figure*}[tb]
\begin{center}
\centering{\epsfig{file=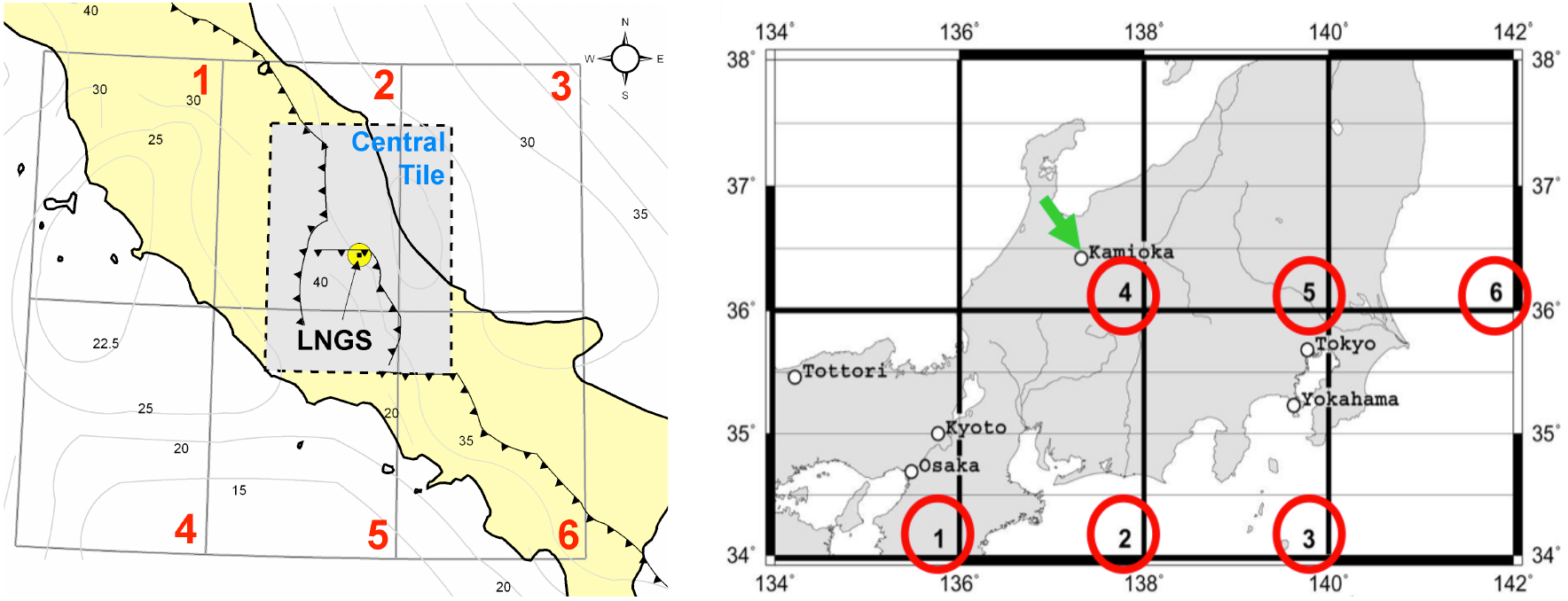,scale=0.9}}
\caption{The map of six 2$^{\circ}$ x 2$^{\circ}$ tiles from which the LOC geo-neutrino signal (see Table~\ref{tab:signalExp}) is calculated for the Borexino (left, from \cite{fabio1}) and KamLAND (right, from \cite{FabioGeoscience}) sites.}
\label{Fig:KBtil}
\end{center}
\end{figure*}

The LOC contributions are taken from~\cite{fiorentini2012}. The calculations are based on six 2$^{\circ} \times 2^{\circ}$ tiles around the detector, as shown in Fig.~\ref{Fig:KBtil}. The LOC contribution in Borexino, based on a detailed geological study of the LNGS area from~\cite{coltorti}, is low, since the area is dominated by dolomitic rock poor in HPE. The LOC contribution in KamLAND is almost double, since the crustal rocks around the site are rich in HPE~\cite{enomoto, fiorentini2005}.

The ROC contributions shown in Table~\ref{tab:signalExp} are taken from~\cite{huang}. This recent crustal model uses as input several geophysical measurements (seismology, gravitometry) and geochemical data as the average compositions of the continental crust~\cite{rudnick} and of the oceanic crust~\cite{white}, as well as several geochemical compilations of deep crustal rocks. The calculated errors are asymmetric due to the log-normal distributions of HPE elements in rock samples. The authors of~\cite{huang} estimate for the first time the geo-neutrino signal from the Continental Lithospheric Mantle (CLM), a relatively thin, rigid portion of the mantle which is a part of the lithosphere (see also Sec.~\ref{Sec:Earth}). 

The mantle contribution to the geo-neutrino signal is associated with a large uncertainty. The estimation of the mass of HPE in the mantle is model dependent. The relatively well known mass of HPE elements in the crust has to be subtracted from the total HPE mass predicted by a specific BSE model. Since there are several categories of BSE models (see Sec.~\ref{Sec:Earth}), the estimates of the mass of HPE in the mantle (and thus of the radiogenic heat from the mantle) varies by a factor of about 8~\cite{sramek}. In addition, the geo-neutrino signal prediction depends on the distribution of HPE in the mantle, which is unknown. As it was described in Sec.~\ref{Sec:Earth}, there are indications of compositional inhomogeneities in the mantle but this is not proved and several authors prefer a mantle with homogeneous composition.  Extremes of the expected mantle geo-neutrino signal with a fixed HPE mass can be defined~\cite{ sramek, fiorentini2012}:
\begin{itemize}
\item{{\it Homogeneous mantle:} the case when the HPE are distributed homogeneously in the mantle corresponds to the {\it maximal, high geo-neutrino signal}. }
\item{{\it Sunken layer:} the case when the HPE are concentrated in a limited volume close to the core-mantle boundary corresponds to the {\it minimal, low geo-neutrino signal}. }
\item{{\it Depleted Mantle + Enriched Layer (DM + EL):}  This is a model of a layered mantle, with the two reservoirs (DM and EL) in which the HPE are distributed homogeneously. The total mass of HPE in the DM + EL corresponds to a chosen BSE model. There are estimates of the composition of the upper mantle (DM), from which the oceanic crust (composed of Mid-Ocean Ridge Basalts, MORB) has been differentiated~\cite{AMCD, SS, WH}. Since in the process of differentiation the HPE are rather concentrated in the liquid part, the residual mantle remains depleted in HPE. The measured  MORB compositions indicate that their source must be in fact depleted in HPE with respect to the rest of the mantle. The mass fraction of the EL is not well defined and in the calculations of \v{S}r\'amek et al.~\cite{sramek} a 427\,km thick EL placed above the core-mantle boundary has been used.}

\end{itemize}

An example of the estimation of the mantle signal for Borexino and KamLAND, given in Table~\ref{tab:signalExp}, is taken from~\cite{huang}.

\begin{table}
\begin{center}
\begin{tabular}{|l|l|l|}
\hline
                                	                     & Borexino  		& KamLAND \\
                                                             & [TNU] & [TNU] \\
 \hline
LOC~\cite{fiorentini2012}                   &     $9.7 \pm 1.3$                   & $17.7 \pm 1.4$     \\
ROC~\cite{huang}                                  &      $13.7^{+2.8}_{-2.3}$     &  $7.3^{+1.5}_{-1.2}$    \\
\hline
Total crust:                                          & $23.4^{+3.1}_{-2.6} $ &  $25.0^{+2.1}_{-1.8}$\\
\hline
CLM~\cite{huang}                                   &      $2.2^{+3.1}_{-1.3}$                  &  $1.6^{+2.2}_{-1.0}$       \\
Mantle~\cite{huang}                               &             8.7           & 8.8     \\  
 \hline
Total  &         $34.3^{+4.4}_{-2.9} $              &   $35.4^{+3.0}_{-2.1} $    \\
\hline
 \end{tabular}
\caption{Expected geo-neutrino signal in Borexino and KamLAND. Details in text.} 
\label{tab:signalExp}
\end{center}
\end{table}

\section{Current experiments }
\label{Sec:experiments}

At the moment, there are only two experiments measuring the geo-neutrinos signals: KamLAND~\cite{KamLANDDet, KamLANDCalib} in the Kamioka mine in Japan and Borexino~\cite{Alimonti1,Alimonti2,Back} at Gran Sasso National Laboratory in central Italy. Both experiments are based on large volume spherical detectors filled with 287\,ton and 1\,kton, respectively, of  liquid scintillator. They both are placed in underground laboratories in order to reduce the cosmic ray fluxes:  a comparative list of detectors' main features is reported in Table \ref{tab:BXKL} .

\begin{table*}
\begin{center}
\begin{tabular}{|l|ll|}
\hline
 ~~                                	                     & Borexino  		&\vline~ KamLAND \\
 \hline
 Depth\dotfill                                          	  & 3600~m.w.e~ ($\phi _{\mu}$=1.2 $m^{-2}h^{-1}$ ) 	&\vline~2700~m.w.e~ ($\phi _{\mu}$=5.4 $m^{-2}h^{-1}$) \\
 Scintillator mass\dotfill                          & 278 ton (PC+1.5g/l PPO)		&\vline~1 kt  (80\% dodec.+20\% PC+1.4g/l PPO)\\
 Inner Detector\dotfill                              & 13\,m sphere, 2212 8'' PMT's	&\vline~18\,m sphere, 1325 17''+554 20'' PMT's  \\
 Outer detector \dotfill                             & 2.4 kt HP water + 208 8'' PMT's  	&\vline~3.2 kt HP water + 225 20'' PMT's \\
 Energy resolution\dotfill                        & 5\% at 1\,MeV 		&\vline~6.4\% at 1\,MeV \\
 Vertex resolution\dotfill                         & 11\,cm at 1\,MeV 		&\vline~12\,cm at 1\,MeV \\
 Reactors mean distance\dotfill            & $\sim$1170\,km		&\vline~$\sim$180\,km\\

 \hline
 \end{tabular}
\caption{Main characteristics of the Borexino and KamLAND detectors.} 
\label{tab:BXKL}
\end{center}
\end{table*}

\subsection{KamLAND}
\label{Sec:KamLand}

The KAMioka Liquid scintillator ANtineutrino Detector (KamLAND) was built, starting from 1999, in a horizontal mine in the Japanese Alps at a depth of 2700 meters water equivalent (m.w.e.). It aimed to a broad experimental program ranging from particle physics to astrophysics and geophysics. 

\begin{figure}[tb]
\begin{center}
\centering{\epsfig{file=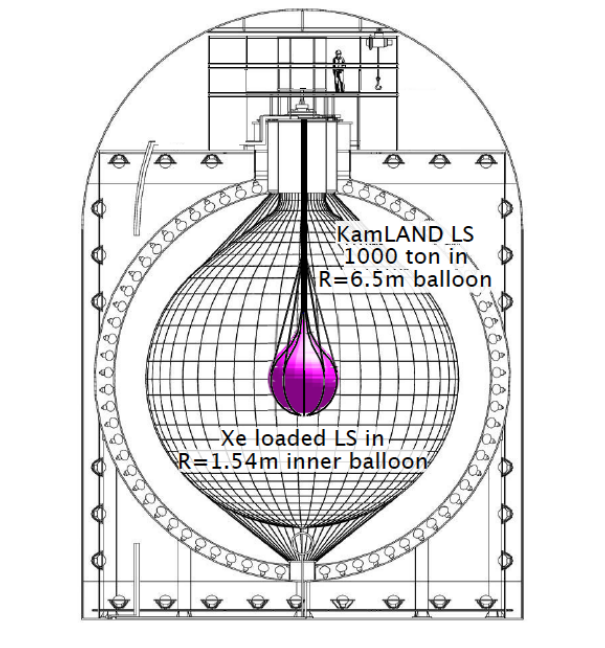,scale=1.4}}
\caption{Schematic view of the KamLAND detector.}
 \label{Fig:KamZen}
\end{center}
\end{figure}

The heart of the detector is a 1\,kton of highly purified liquid scintillator, made of  80\% dodecane, 20\% pseudocumene, and 1.36 $\pm$ 0.03\,g/l of  2,5-Diphenyloxazole (PPO). It is characterized by a high scintillation yield, high light transparency and a fast decay time, all essential requirements for good energy and spatial resolutions.
The scintillator is enclosed in a 13\,m spherical nylon balloon, suspended in a non-scintillating mineral oil by means of Kevlar ropes and contained inside a 9\,m-radius stainless-steel tank (see Fig.~\ref{Fig:KamZen}). An array of 1325 of 17'' PMTs and 554 of 20'' PMTs (Inner Detector) is mounted inside the stainless-steel vessel viewing the center of the  scintillator sphere and providing a 34\% solid angle coverage. The containment sphere is surrounded by a 3.2 kton cylindrical water Cherenkov Outer Detector that shields the external background and acts as an active cosmic-ray veto.

The  KamLAND detector is  exposed to a very large flux of low-energy antineutrinos coming from the nuclear reactor plants.
Prior to the earthquake and tsunami of March 2011, one-third of all Japanese electrical power (which is equivalent to 130\,GW thermal power) was provided by nuclear reactors. The fission reactors release about 10$^{20}$ $\bar{\nu}_e$ GW$^{-1}$s$^{-1}$ that mainly come from the $\beta$-decays of the fission products of $^{ 235}$U,$^{ 238}$U, $^{239}$Pu, and $^{241}$Pu, used as fuels in reactor cores. The mean distance of reactors from KamLAND is $\sim$180\,km. Since 2002, KamLAND is detecting hundreds of $\bar{\nu}_e$ interactions per year.

The first success of the collaboration, a milestone in the neutrino and particle physics, was to provide a direct evidence of  the neutrino flavor oscillation by observing the reactor $\bar{\nu}_e$ disappearance~\cite{eguchi2003} and the energy spectral distortion as a function of the distance to $\bar{\nu}_e$-energy ratio~\cite{araki2005}. The measured oscillation parameters, $\Delta m^{2}_{21}$ and $\tan^{2}(\theta_{12})$, were found, under the hypothesis of CPT invariance, in agreement with the Large Mixing Angle (LMA) solution to the solar neutrino problem, and the precision of the $\Delta m^{2}_{21}$ was greatly improved. In the following years, the oscillation parameters were measured with increasing precision~\cite{abe2008}. 

KamLAND was the first experiment to perform experimental investigation of geo-neutrinos in 2005~\cite{araki2005b}.
An updated geo-neutrino analysis was released in 2008~\cite{abe2008}. An extensive liquid-scintillator purification campaign to improve its radio-purity took place in years 2007 - 2009. Consequently, a new geo-neutrino observation at 99.997\% C.L. was achieved in 2011 with an improved signal-to-background ratio~\cite{gando2011}.  Recently, after the earthquake and the consequent Fukushima nuclear accident occurred in March 2011, all Japanese nuclear reactors were temporarily switched off for a safety review.  Such situation allowed for a reactor on-off study of backgrounds and also yielded an improved sensitivity for $\bar{\nu}_e$ produced by other sources, like geo-neutrinos.  A new result on geo-neutrinos has been released recently in March 2013~\cite{gando2013b}.

In September 2011, the KamLAND-Zen $\nu$-less double beta-decay search was launched. A $\beta\beta$ source, made up by 13\,ton of Xe-loaded liquid scintillator was suspended inside a 3.08\,m diameter inner balloon placed at the center of the detector (see Fig.~\ref{Fig:KamZen}). A new lower limit for the $\nu$-less double-beta decay half life was published in 2013~\cite{gando2013}.

\subsection{Borexino}
\label{Borexino}

The Borexino detector was built starting from 1996 in the underground hall C of the Laboratori Nazionali del Gran Sasso in Italy, with the main scientific goal to measure in real-time the low-energy solar neutrinos. Neutrinos are even more tricky to be detected than antineutrinos. In a liquid scintillator, $\bar{\nu}_e$'s give a clean delayed-coincidence tag which helps to reject backgrounds, see Sec.~\ref{Sec:detection}.
Neutrinos, instead, are detected through their scattering off electrons which does not provide any coincidence tag. The signal is virtually indistinguishable from any background giving a $\beta$/$\gamma$ decays in the same energy range.
For this reason, an extreme radio-purity of the scintillator, a mixture of pseudocumene and PPO as fluor at a concentration of 1.5\,g/l, was an essential pre-requisite for the success of Borexino.

\begin{figure}[tb]
\begin{center}
\centering{\epsfig{file=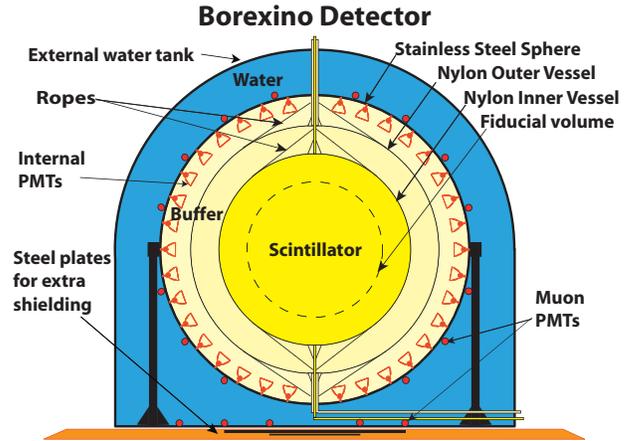,scale=0.3}}
\caption{Schematic diagram of the Borexino detector.}
\label{Fig:Bdet}
\end{center}
\end{figure}

For almost 20 years the Borexino collaboration has been addressing this goal by developing advanced purification techniques for scintillator, water, and nitrogen and  by exploiting innovative cleaning systems for each of the carefully selected materials. A  prototype of the Borexino detector, the Counting Test Facility (CTF)~\cite{CTF1-1,CTF1-2} was built to prove the purification effectiveness.
The conceptual design of Borexino is based on the principle of graded shielding demonstrated in Fig.~\ref{Fig:Bdet}. A set of concentric shells of increasing radio-purity moving inwards surrounds the inner scintillator core. The core is made of $\sim$280 ton of scintillator, contained in a 125 $\mu$m thick nylon Inner Vessel (IV) with a radius of 4.25\,m and shielded from external radiation by 890\,ton of inactive buffer fluid.  Both the active and inactive layers are contained in a 13.7\,m diameter Stainless Steel Sphere (SSS) equipped with 2212 8'' PMTs (Inner Detector). A cylindrical dome with diameter of 18\,m and height of 16.9\,m encloses the SSS. It is filled with 2.4\, kton of ultra-pure water viewed by 208 PMT's defining the Outer Detector. The external water serves both as a passive shield against external background sources, mostly neutrons and gammas, and also as an active Cherenkov veto system tagging the residual cosmic muons crossing the detector.

After several years of construction, the data taking started in May 2007, providing immediately evidence of the unprecedented scintillator radio-purity. Borexino was the first experiment to measure in real time low-energy solar neutrinos below 1\,MeV, namely the $^{7}$Be-neutrinos~\cite{BXbe7arp,BXbe7192}. In May 2010, the Borexino Phase 1 data taking period was concluded. Its main scientific goal, the precision $^{7}$Be-$\nu$ measurement has been achieved~\cite{BXbe7prec} and the absence of the day-night asymmetry of its interaction rate was observed~\cite{BXbe7asim}. In addition, other major goals were reached, as the first observation of the $pep$-$\nu$ and the strongest limit on the CNO-$\nu$~\cite{BXpep}, the measurement of $^{8}$B-$\nu$ rate with a 3\,MeV energy threshold~\cite{BXb8}, and in 2010, the first observation of geo-neutrinos with high statistical significance at 99.997\% C.L.~\cite{BXgeo1}.

In 2010-2011 six purification campaigns were performed to further improve the detector performances and in October 2011, the Borexino Phase 2 data taking period was started. A new result on geo-neutrinos has been released in March 2013~\cite{BXgeo2}. Borexino continues in a rich solar neutrino program, including two even more challenging targets: $pp$ and possibly CNO neutrinos. In parallel, the Borexino detector will be used in the SOX project, a short baseline experiment, aiming at investigation of the sterile-neutrino hypothesis~\cite{BXsox}.

\section{Geo-neutrino analysis}
\label{Sec:analysis}

\subsection{The geo-neutrino detection}
\label{Sec:detection}

The hydrogen nuclei that are copiously present in hydrocarbon (C$_{n}$H$_{2n}$) liquid scintillator detectors act as target for electron antineutrinos in the inverse beta decay reaction shown in Eq.~\ref{Eq:InvBeta}. In this process, a positron and a neutron are emitted as reaction products. The positron promptly comes to rest and annihilates emitting two 511\,keV $\gamma$-rays, yielding a {\it prompt event}, with a visible energy $E_{prompt}$, directly correlated with the incident antineutrino energy $E_{\bar{\nu}_e}$: 
\begin{equation}
E_{prompt}= E_{\bar{\nu}_e}- 0.784 MeV.  
\label{Epro}
\end{equation}
The emitted neutron keeps initially the information about the ${\bar{\nu}_e}$ direction, but, unfortunately, the neutron is typically captured on protons only after a quite long thermalization time ($\tau$ = 200 - 250\,$\mu$s, depending on scintillator). During this time, the directionality memory is lost in many scattering collisions. When the thermalized neutron is captured on proton, it gives a typical 2.22\,MeV de-excitation $\gamma$-ray, which provides a coincident {\it delayed event}.  The pairs of time and spatial coincidences between the prompt and the delayed signals offer a clean signature of $\bar{\nu}_e$ interactions, very different  from the $\nu_e$ scattering process used in the neutrino detection.

\subsection{Background sources}

The coincidence tag used in the electron antineutrino detection is a very powerful tool in background suppression.  The main antineutrino background in the geo-neutrino measurements results from nuclear power plants, while negligible signals are due to atmospheric and relic supernova $\bar{\nu}_e$. Other, non-antineutrino background sources can arise from intrinsic detector contamination's, from random coincidences of non-correlated events, and from cosmogenic sources, mostly residual muons. An overview of the main background sources in the Borexino and KamLAND geo-neutrino measurements is presented in Table~\ref{tab:BXKL2}. 

A careful analysis of the expected reactor $\bar{\nu}_e$ rate at a given experimental site is crucial. The determination of the expected signal from reactor $\bar{\nu}_e$'s requires the collection of the detailed information on the time profiles of the thermal power and nuclear fuel composition for all the reactors, especially for the nearby ones. The Borexino and KamLAND collaborations are in strict contact with the International Agency of Atomic Energy (I.A.E.A.) and the Consortium of Japanese Electric Power Companies, respectively.

\begin{figure}[tb]
\begin{center}
\centering{\epsfig{file=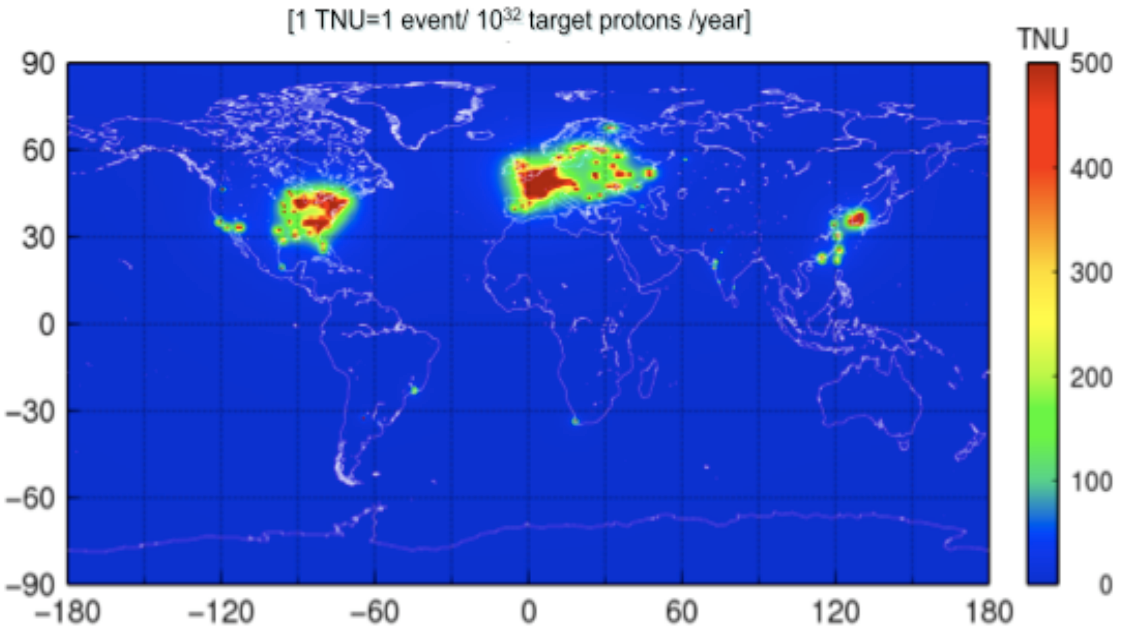,scale=0.8}}
\caption{Reactor ${\bar{\nu}_e}$ signal (expressed in TNU) in the world as in the middle of 2012, calculated in~\cite{ricci}.}
\label{Fig:fluxba}
\end{center}
\end{figure}
A new recalculation~\cite{mueller,huber2013} of the $\bar{\nu}_e$ spectra per fission of $^{ 235}$U,$^{ 238}$U, $^{239}$Pu, and $^{241}$Pu isotopes predicted a $\sim$3\% flux increase relative to the previous calculations. As a consequence, all past experiments at short-baselines appear now to have seen fewer $\bar{\nu}_e$ than expected and this problem was named the Reactor Neutrino Anomaly~\cite{mention}. It has been speculated that it may be due to some not properly understood systematics but  in principle an  oscillation into an hypothetical heavy sterile neutrino state with $\Delta m^{2} \sim$1eV$^{2}$ could explain this anomaly. In the KamLAND analysis, the cross section per fission for each reactor was normalized to the experimental fluxes measured by Bugey-4~\cite{mention}. The Borexino analysis is not affected by this effect since the absolute reactor antineutrino signal was left as a free parameter in the fitting procedure and the spectral shape of the new parametrization is not significantly different up to 7.5\,MeV from the previous ones.

The expected reactor $\bar{\nu}_e$ signal in the world~\cite{ricci} is shown in Fig.~\ref{Fig:fluxba}: it refers to the middle of 2012 when the Japanese nuclear power plants were switched off. The red spot close to Japan is due to the Korean reactors. The world average nuclear energy production is of the order of 1 TW, a 2\% of the Earth surface heat flux. There are no nuclear power plants in Italy, and the reactor $\bar{\nu}_e$ flux in Borexino is a factor of 4-5 lower then in the KamLAND site during normal operating condition.

\begin{table*}
\begin{center}
\begin{tabular}{|l|ll|}
\hline
 ~~                                	    & Borexino  		&\vline~ KamLAND \\
 \hline
Period  & Dec 07 - Aug 12  &\vline~Mar 02 - Nov 12\\
Exposure (proton $ \cdot$ year) 	& (3.69 $\pm$ 0.16) 10$^{31}$ &\vline~(4.9 $\pm$ 0.1) 10$^{32}$\\
\hline
Reactor-$\bar{\nu}_e$ events (no osc.)  & 60.4 $\pm$ 4.1 &\vline~3564 $\pm$ 145\\
\hline
$^{13}$C($\alpha$, n)$^{16}$O~events & 0.13 $\pm$ 0.01 &\vline~207.1 $\pm$ 26.3\\
$^{9}$Li - $^{8}$He~events  & 0.25 $\pm$ 0.18 &\vline~31.6 $\pm$ 1.9\\
Accidental events & 0.206 $\pm$ 0.004 &\vline~125.5 $\pm$ 0.1\\
\hline
Total non-$\bar{\nu}_e$ backgrounds & 0.70 $\pm$ 0.18 &\vline~364.1 $\pm$ 30.5\\
 \hline
 \end{tabular}
\caption{The most important backgrounds in geo-neutrino measurements of Borexino~\cite{BXgeo2} and KamLAND~\cite{gando2013b}.} 
\label{tab:BXKL2}
\end{center}
\end{table*}

A typical rate of $\sim$5 and $\sim$21 geo-$\nu$ events/year with 100\% efficiency is expected in the Borexino and KamLAND detector, for a 4 m and 6 m fiducial volume cut, respectively. This signal is very faint and also the non-$\bar{\nu}_e$-induced backgrounds have to be incredibly small.
Random coincidences and ($\alpha$, n) reactions in which $\alpha$'s are mostly due to the $^{210}$Po decay (belonging to the $^{238}$U chain) can mimic the reaction of interest. The $\alpha$-background was particularly high for the KamLAND detector at the beginning of data taking ($\sim$10$^{3}$ cpd/ton) but it has been successfully reduced by a factor 20 thanks to the 2007 - 2009 purification campaigns.
Backgrounds faking $\bar{\nu}_e$ interactions could also arise from cosmic muons and muon induced neutrons and unstable nuclides like $^{9}$Li and $^{8}$He having an $\beta$+neutron decay branch. Very helpful to this respect is the rock overlay of 2700 m.w.e for the KamLAND and 3600 m.w.e for the Borexino experimental site, reducing this background by a factor up to 10$^{6}$. A veto applied after each muon crossing the Outer and/or the Inner Detectors, makes this background almost negligible.

\subsection{Current experimental results}
\label{Sec:results}

Both Borexino~\cite{BXgeo2} and KamLAND~\cite{gando2013b} collaborations released new geo-neutrino results in March 2013 and we describe them in more detail below. The corresponding geo-neutrino signals and signal-to-background ratios are shown in Table~\ref{tab:BXKL2}.

\begin{table*}
\begin{center}
\begin{tabular}{|l|ll|}
\hline
 ~~                                	    & Borexino  		&\vline~ KamLAND \\
 \hline
Period  & Dec 07- Aug 12  &\vline~Mar 02- Nov 12\\
Exposure (proton $ \cdot$ year) 	& (3.69 $\pm$ 0.16) 10$^{31}$ &\vline~(4.9 $\pm$ 0.1) 10$^{32}$\\
\hline
Geo-$\nu$ events 	& 14.3 $\pm$ 4.4 &\vline~116 $^{+28}_{-27}$\\
Geo-$\nu$ signal [TNU]   & 38.8 $\pm$ 12 &\vline~30 $\pm$ 7\\
Geo-$\nu$ flux (oscill.) $[ \cdot 10^{6}$\,cm$^{-2}$s$^{-1}$]   & 4.4 $\pm$ 1.4 &\vline~3.4 $\pm$ 0.8\\
\hline
Geo-$\nu$ signal/(not-oscill. anti-$\nu$ background) & 0.23 &\vline~0.032\\
Geo-$\nu$ signal/(non anti-$\nu$ background) & 20.4 &\vline~0.32\\
 \hline
 \end{tabular}
\caption{Measured geo-neutrino signal in Borexino~\cite{BXgeo2} and KamLAND~\cite{gando2013b}.} 
\label{tab:BXKL3}
\end{center}
\end{table*}

The KamLAND result is based on a total live-time of 2991 days, collected between March 2002 and November 2012. In this 10-year time window the backgrounds and detector conditions have changed.  
After the April 2011 earthquake the Japanese nuclear energy production was strongly reduced and in particular in the April to June 2012 months all the Japanese nuclear reactors were switched off with the only exception of the Tomary plant which is in any case quite far ($\sim$600 km) from the KamLAND site. This reactor-off statistics was extremely helpful to check all the other backgrounds and it is included in the present data sample even if with a reduced Fiducial Volume (FV). In fact, because of the contemporary presence of the Inner Balloon containing the Xe loaded scintillator at the detector center, the central portion of the detector was not included in the analysis. 

\begin{figure*}[tb]
\begin{center}
\centering{\epsfig{file=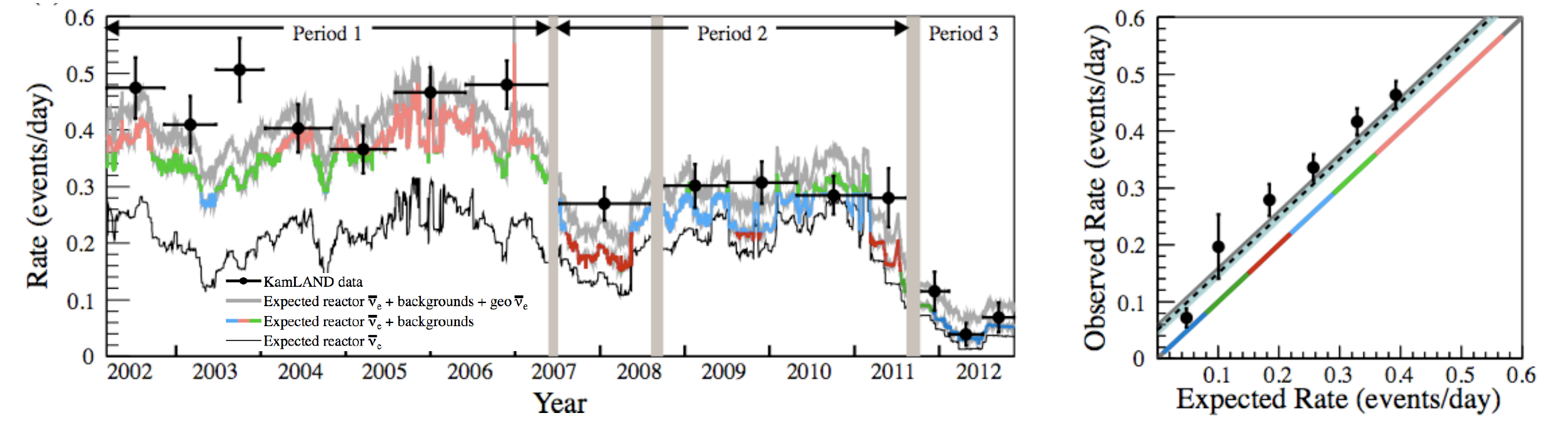,scale=0.7}}
\begin{minipage}[t]{16.5 cm}
\caption{Left: Event rate in the KamLAND detector as a function of time in the 0.9-2.6 energy window. Right: The excess of events with respect to the expected background rate is constant in time and attributed to the geo-$\nu$ signal. Taken from~\cite{gando2013b}.}
\label{Fig:KLrate}
\end{minipage}
\end{center}
\end{figure*}

The $\bar{\nu}_e$ event rate in the KamLAND detector and in the energy window 0.9 - 2.6\,MeV as a function of time is shown in Fig.~\ref{Fig:KLrate}-left. The measured excess of events with respect to the background expectations is constant in time, as highlighted in Fig.~\ref{Fig:KLrate}-right, and is attributed to the geo-neutrino signal.

\begin{figure*}[tb]
\begin{center}
\centering{\epsfig{file=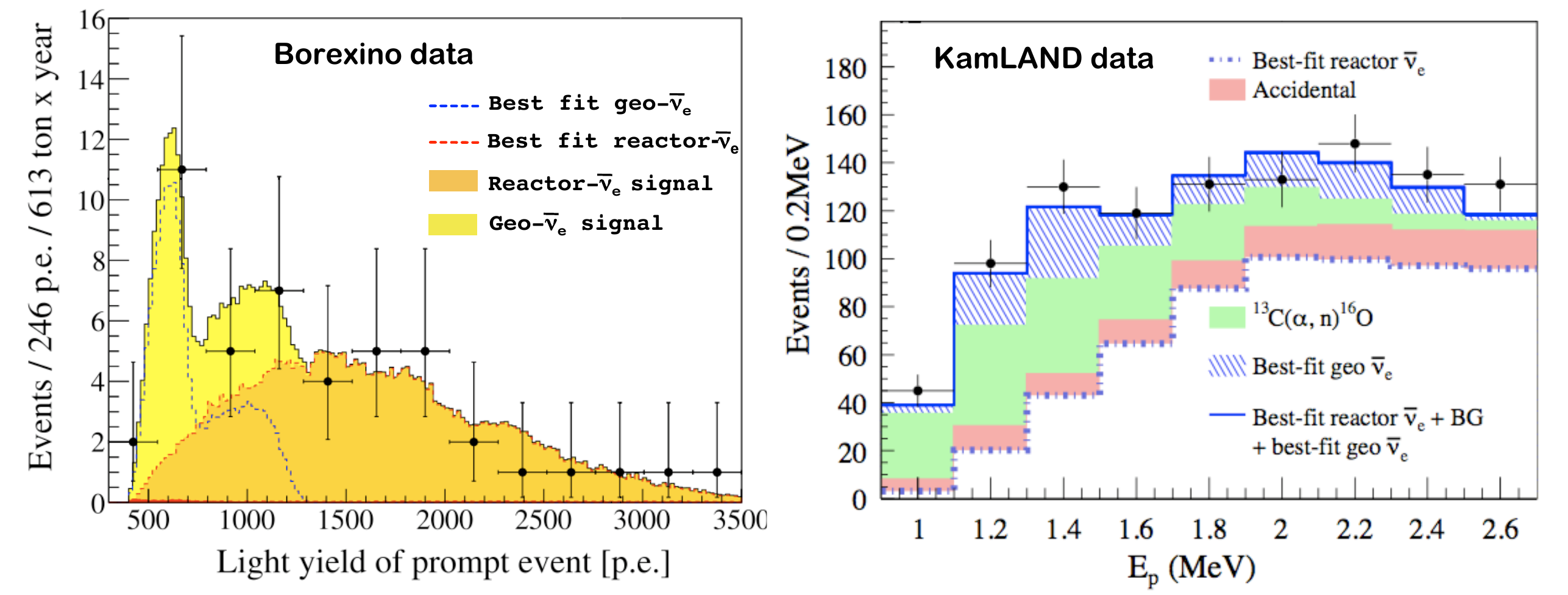,scale=0.67}}
\begin{minipage}[t]{16.5 cm}
\caption{Prompt event energy spectrum measured in Borexino (left) and in KamLAND (right). The Borexino collaboration quotes the prompt event energy as total number of photoelectrons detected by the PMTs, the conversion factor being approximately 500\,p.e./1\,MeV.}
\label{Fig:KBspe}
\end{minipage}
\end{center}
\end{figure*}

To extract the neutrino oscillation parameters and the geo-neutrino fluxes, the $\bar{\nu}_e$ candidates are analyzed with an unbinned maximum likelihood method incorporating the measured event rates, the energy spectra of prompt candidates and their time variations. The total energy spectrum of prompt events and the fit results are shown in Fig.~\ref{Fig:KBspe}-right. By assuming a chondritic Th/U mass ratio of 3.9, the fit results in $116^{+28}_{ -27}$ geo-neutrino events, corresponding to a total oscillated flux of $3.4^{+0.8}_{ -0.8} \cdot10^{6}$\,cm$^{-2}$ s$^{-1}$. 
It is easy to demonstrate that given the geo-neutrino energy spectrum, the chondritic mass ratio, and the inverse beta decay cross section, a simple conversion factor exists between the fluxes and the TNU units: 1 TNU = 0.113~$\cdot$ 10$^{6}$~$\bar{\nu}_e$ cm$^{-2}$s$^{-1}$. By taking this factor we could translate the KamLAND result to $(30 \pm 7)$\,TNU.

While the precision of the KamLAND result is mostly affected by the systematic uncertainties arising from the sizeable backgrounds, the extremely low background together with the smaller fiducial mass (see Tables~\ref{tab:BXKL2} and \ref{tab:BXKL3}) makes the statistical error largely predominant in the Borexino measurement.

The Borexino result, shown in Fig.~\ref{Fig:KBspe}-left, refers to the statistics collected from December 2007 to August 2012. The levels of background affecting the geo-$\nu$ analysis were almost constant during the whole data taking, the only difference being an increased radon contamination during the test phases of the purification campaigns. These data periods are not included in the solar neutrino analysis but can be used in the geo-neutrino analysis. A devoted data selection cuts were adopted to make the increased background level not significant, in particular, an event pulse-shape analysis and an increased energy threshold have been applied for delayed candidates.

The Borexino collaboration selected 46 antineutrino candidates (Fig.~\ref{Fig:KBspe}-left), among which  33.3 $\pm$ 2.4 events were expected from nuclear reactors and 0.70 $\pm$ 0.18 from the non-$\bar{\nu}_e$ backgrounds. An unbinned maximal likelihood fit of the light-yield spectrum of prompt candidates was performed, with the Th/U mass ratio fixed to the chondritic value of 3.9, and with the number of events from reactor antineutrinos left as a free parameter. As a result, the number of observed geo-neutrino events is 14.3 $\pm$ 4.4 in (3.69 $\pm$ 0.16)~$\cdot$~10$^{31}$ proton $\cdot$ year exposure. 
This signal corresponds to $\bar{\nu}_e$ fluxes from U and Th chains, respectively, of $\phi$(U) = (2.4 $\pm$ 0.7)~$\cdot$~10$^{6}$\,cm$^{-2}$s$^{-1}$ and $\phi$(Th) = (2.0 $\pm$ 0.6)~$\cdot$~10$^{6}$\,cm$^{-2}$s$^{-1}$ and to a total measured normalized rate of (38.8 $\pm$ 12)\,TNU. 

The measured geo-neutrino signals reported in Table~\ref{tab:BXKL3} can be compared with the expectations reported in Table~\ref{tab:signalExp}. The two experiments placed very far form each other have presently measured the geo-neutrino signal with a high statistical significance (at $\sim$$4.8\sigma$ C.L.) and in a good agreement with the geological expectations. This is an extremely important point since it is confirming both that
the geological models are working properly and that the geo-neutrinos are a reliable tools to investigate the Earth structure and composition.

\subsection{Geological implications}
\label{Sec:Heat}

\begin{figure}[tb]
\begin{center}
\centering{\epsfig{file=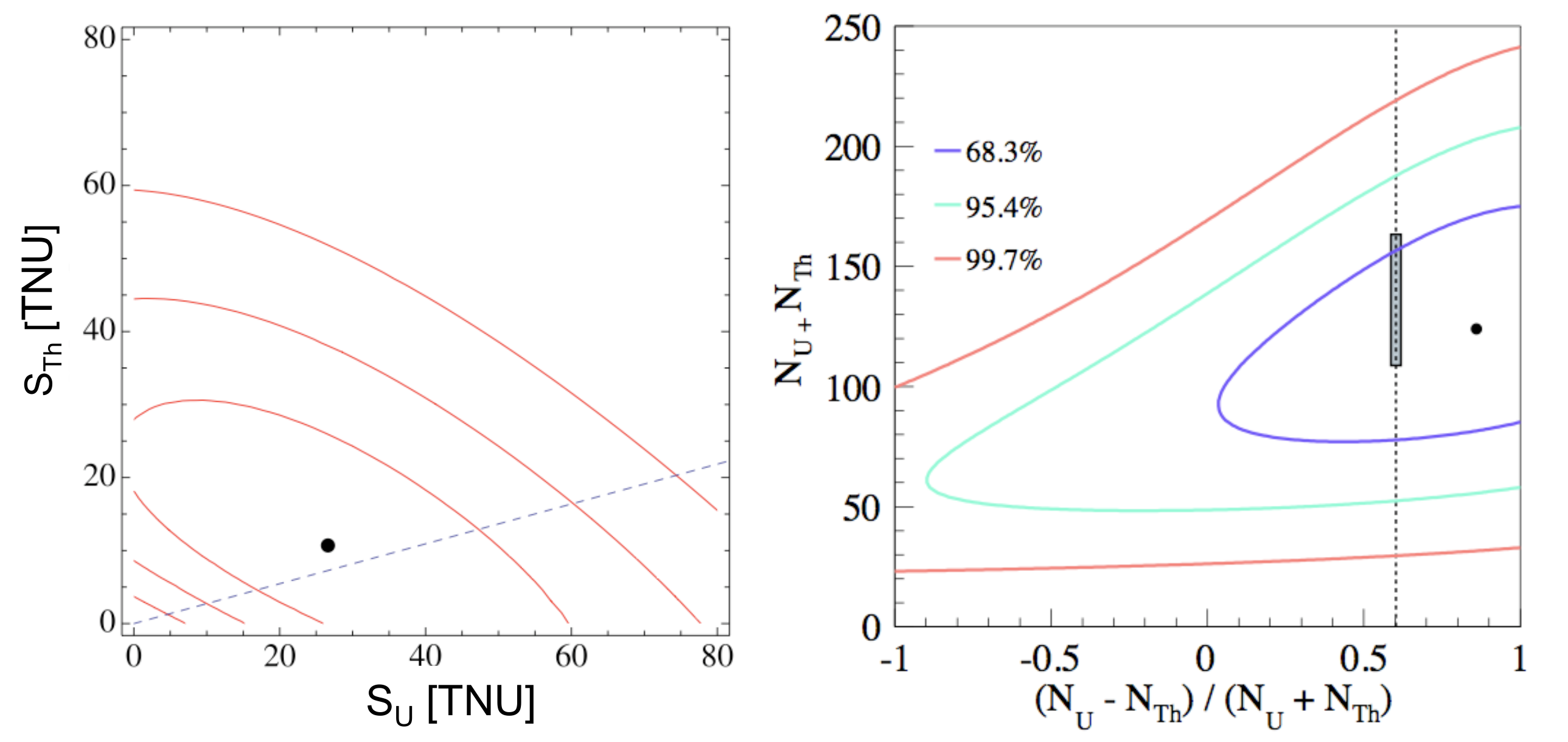,scale=0.35}}
\caption{Left: The 68.3, 95.4, and 99.7\% countour plots of the Th versus U signal, expressed in TNU units, in the Borexino geo-neutrino analysis~\cite{BXgeo2}; the dashed blue line is the expectation for a chondritic Th/U mass ratio of 3.9. Right: the same confidence level contours are shown for the KamLAND analysis~\cite{gando2013b}, expressed in number of total events versus the normalized difference of the number of events from U and Th. The vertical dashed line represents the chondritic ratio of 3.9 while the shadowed area on this line is the prediction of the BSE model from~\cite{McDonoughSun}. }
\label{Fig:UTh}
\end{center}
\end{figure}

In the standard geo-neutrino analysis, the Th/U bulk mass ratio has been assumed to be 3.9, a value of this ratio observed in CI chondritic meteorites and in the solar photosphere, and, a value assumed by the geo-chemical BSE models. However, this value has not yet been experimentally proven for the bulk Earth. The knowledge of this ratio would be of a great importance in a view of testing the geo-chemical models of the Earth formation and evolution. It is, in principle, possible to measure this ratio with geo-neutrinos, exploiting the different end-points of the energy spectra from U and Th chains (see Fig.~\ref{Fig:geonuspe}). A mass ratio of $m$(Th)/$m$(U) = 3.9 corresponds to the signal ratio $S$(U)/$S$(Th) $\sim$3.66. Both KamLAND and Borexino collaborations attempted an analysis in which they tried to extract the individual U and Th contributions by removing the chondritic constrain from the spectral fit. In Fig.~\ref{Fig:UTh}, the confidence-level contours from such analyses are shown for Borexino (left) and for KamLAND (right). Borexino has observed the central value a $S$(U)/$S$(Th) of $\sim$2.5 while KamLAND of $\sim$14.5 but they are not in contradiction since the uncertainties are still very large and the results not at all conclusive. Both the best fit values are compatible at less than 1$\sigma$ level with the chondritic values. 

\begin{figure}[tb]
\begin{center}
\vspace{-2 mm}
\centering{\epsfig{file=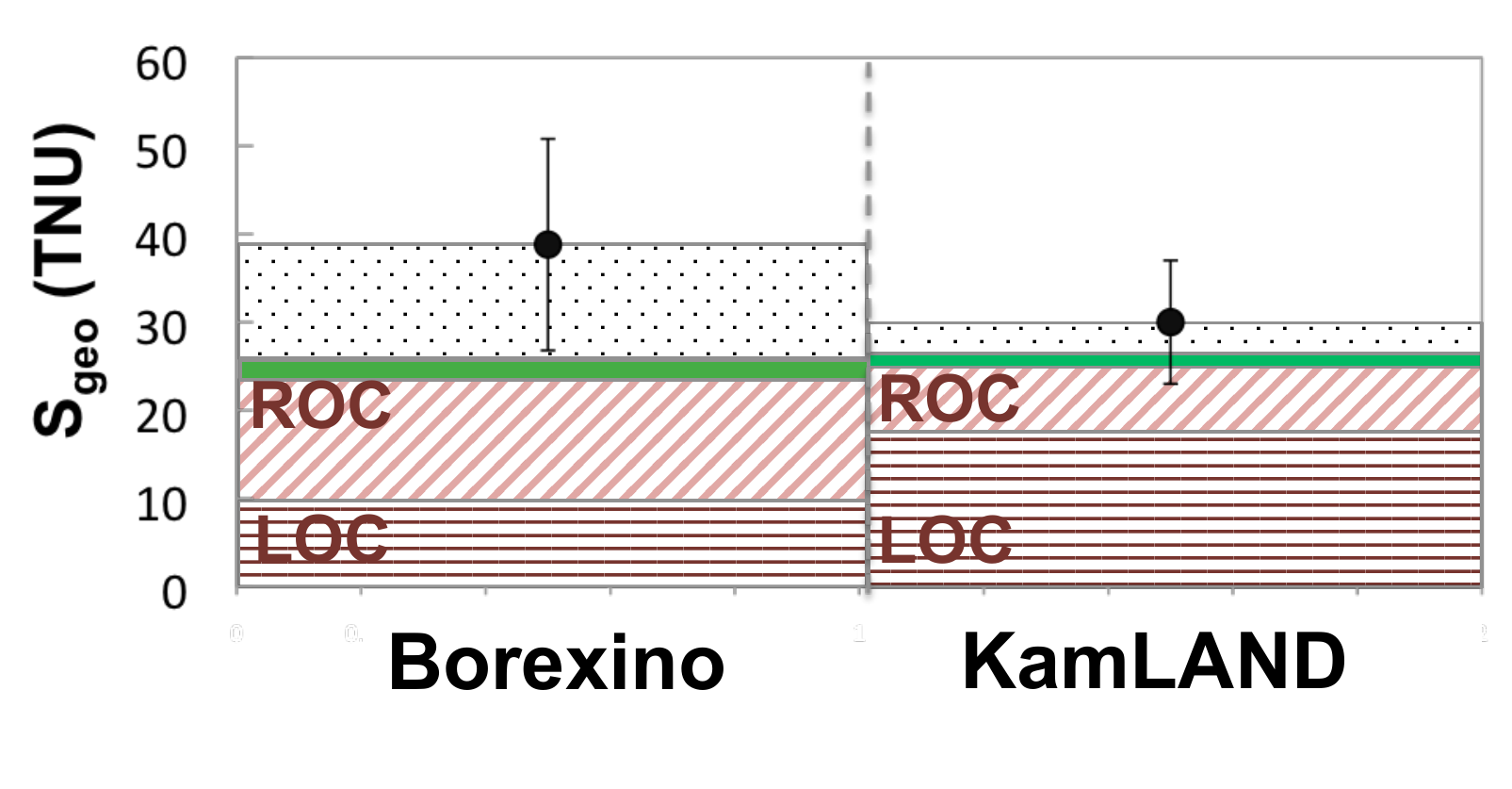,scale=0.55}}
\caption{The measured geo-$\nu$ signal in Borexino and KamLAND compared to the expected fluxes from Table~\ref{tab:signalExp}: area with horizontal stripes = LOC, area with oblique stripes = ROC, green solid area = CLM. The dotted area is the excess of signal which could correspond to the convective mantle contributions. The sum of the CLM and the convective mantle contributions corresponds to the total mantle signal as from Eq.~\ref{Eq:mantleKL} and \ref{Eq:mantleBX}.}
\label{Fig:KBstrati}
\end{center}
\end{figure}

As discussed in Sec.~\ref{Sec:Earth}, the principal goal of geo-neutrino measurements is to determine the HPE abundances in the mantle and from that to extract the strictly connected radiogenic power of the Earth. 
The geo-neutrino fluxes from different reservoirs sum up at a given site, so the mantle contribution can be inferred from the measured signal by subtracting the estimated crustal (LOC + ROC) components (Sec.~\ref{Sec:signal}). Considering the expected crustal signals from Table~\ref{tab:signalExp} and the measured geo-neutrino signals from Table~\ref{tab:BXKL3}, 
such a simple subtraction results in mantle signals measured by KamLAND $S_M^{KL}$ and Borexino $S_M^{BX}$ of:
\begin{equation}
S_{Mantle}^{KL} = (5.0 \pm 7.3)~{\rm TNU}
\label{Eq:mantleKL}
\end{equation}
\begin{equation}
S_{Mantle}^{BX} = (15.4 \pm 12.3)~{\rm TNU}
\label{Eq:mantleBX}
\end{equation}
A graphical representation of the different contributions in the measured signals is shown in Fig.~\ref{Fig:KBstrati}.

The KamLAND result seems to highlight a smaller mantle signal than the Borexino one. Such a result pointing towards mantle inhomogeneities is very interesting from a geological point of view, but the error bars are still too large to get statistically significant conclusions. Indeed, recent models predicting geo-neutrino fluxes from the mantle not spherically symmetric have been presented~\cite{sramek}. They are based on the hypothesis, indicated by the geophysical data, that the Large Low Shear Velocity Provinces (LLSVP) placed at the base of the mantle beneath Africa and the Pacific represent also compositionally distinct areas. In a particular, the TOMO model~\cite{sramek} predicts a mantle signal in Borexino site higher by 2\% than the average mantle signal while a decrease of 8.5\% with respect to the average is expected for KamLAND. We have performed a combined analysis of the Borexino and KamLAND data in the hypothesis of a spherically symmetric mantle or a not homogeneous one as predicted by the TOMO model.

\begin{figure}[tb]
\begin{center}
\vspace{-4 mm}
\hspace{4 mm}
\centering{\epsfig{file=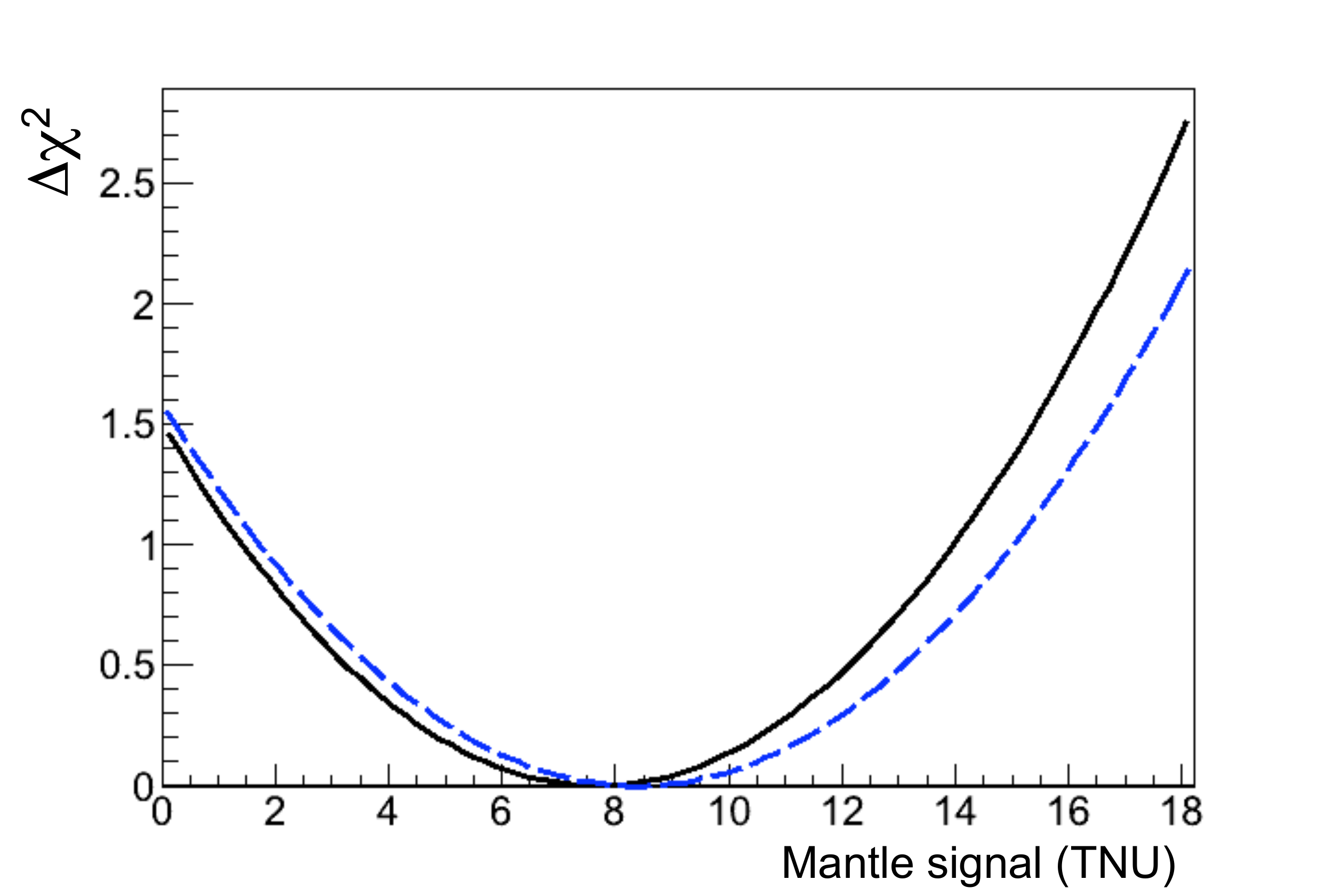,scale=0.33}}
\caption{ $\Delta \chi^{2}$ profile for the mantle signal in the Borexino +  KamLAND combined analysis. The black continuous line assumes a spherically symmetric mantle, while the dashed blue line a non homogeneous TOMO model from~\cite{sramek}.} 
\label{Fig:KBdchi2}
\end{center}
\end{figure}

The $\Delta \chi^{2}$  profiles for both models are shown in Fig.~\ref{Fig:KBdchi2}. For the homogeneous mantle we have obtained the signal $S_{Mantle}^{SYM}$ of 
\begin{equation}
S_{Mantle}^{SYM} = (7.7 \pm 6.2)~{\rm TNU}.
\label{Eq:homo}
\end{equation}
Instead, when the Borexino and KamLAND mantle signals have been constrained to the ratio predicted by the TOMO model, the mean mantle signal $S_{Mantle}^{TOMO}$ results to be
\begin{equation}
S_{Mantle}^{TOMO}= (8.4^{+6.6}_{-6.7})~{\rm TNU}.
\label{Eq:tom}
\end{equation}
There is an indication for a positive mantle signal but only with a small statistical significance  of about 1.5$\sigma$ C.L.
The central values are quite in agreement with the expectation shown in Table~\ref{tab:signalExp}. A slightly higher central value is observed for the TOMO model. We 
stress again the importance of a detailed knowledge of the local crust composition and thickness in order to deduce the signal coming from the mantle from the
measured geo-neutrino fluxes.

\begin{figure}[tb]
\begin{center}
\centering{\epsfig{file=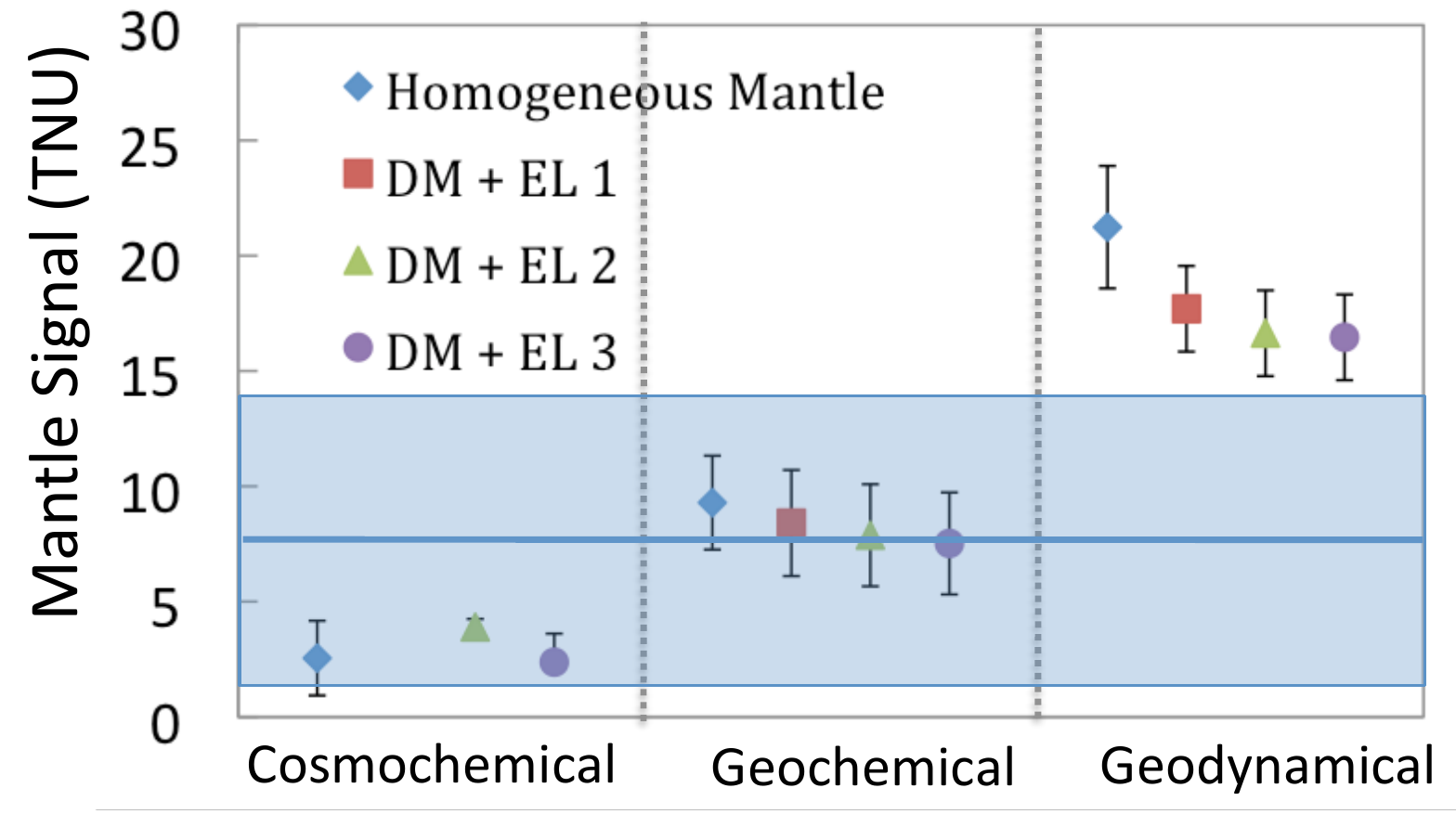,scale=0.55}}
\caption{The measured geo-neutrino signal from the Borexino + KamLAND combined analysis under the assumption of a spherically symmetric mantle (see~Eq.~\ref{Eq:homo}) is compared with the predictions of different Earth's models from~\cite{sramek}. The three DM + EL distributions of the HPE elements in the mantle correspond to the depleted mantle compositions from~\cite{AMCD, SS, WH}, respectively.}
\label{Fig:KBsra}
\end{center}
\end{figure}

In Fig.~\ref{Fig:KBsra}, we compare the measured mantle signal $S_{Mantle}^{SYM}$ from Eq.~\ref{Eq:homo} with the predictions of the three categories of the BSE models according to~\cite{sramek} which we have discussed in Sec.~\ref{Sec:signal}, i.e. the geochemical, cosmochemical, and geodynamical ones. For each BSE model category, four different HPE distributions through the mantle have been considered: a homogeneous model and the three DM + EL models with the three different depleted mantle compositions as in~\cite{AMCD, SS, WH}. All the Earth models are still compatible at 2$\sigma$ level with the measurement, as shown in Fig.~\ref{Fig:KBsra}, even if the present combined analysis slightly disfavors the geodynamical models. We remind that these models are based on the assumption that the radiogenic heat has provided the power to sustain the mantle convection over the whole Earth story. It has been recently understood~\cite{crow} the importance of the water or water vapor embedded in the crust and mantle to decrease the rock viscosity and so the energy supply required to promote the convection. If this is the case the geodynamical models are going to be reconciled with the geochemical ones.

\begin{figure}[tb]
\begin{center}
\centering{\epsfig{file=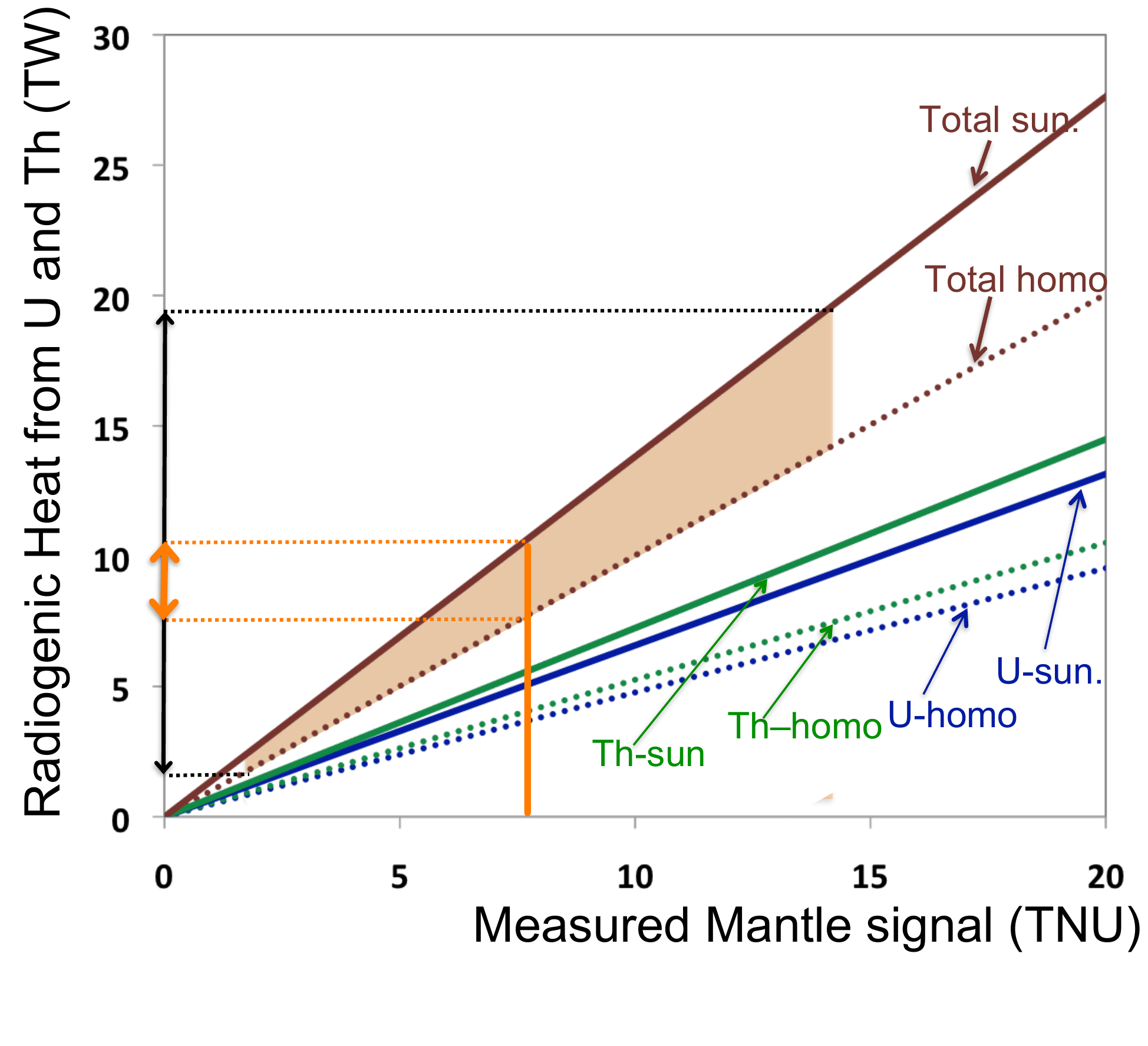,scale=0.45}}
\caption{The mantle radiogenic heat power from U and Th as a function of the measured geo-neutrino signal; the solid lines represent the sunken-layer model, while the dotted lines the homogeneous mantle (see Sec.~\ref{Sec:signal}). The green and the blue lines indicate the individual Th and U contributions, respectively, while the brown lines show the total signal. The measured mantle geo-neutrino signal  $S_{Mantle}^{SYM}$ from a combined Borexino + KamLAND analysis is shown by the vertical solid orange line; the corresponding 1$\sigma$ band is shown by a filled triangular area. The arrows on the vertical $y$-axis indicate the radiogenic heat corresponding to the best fit geo-neutrino signal. Details in text.}
\label{Fig:KBTW}
\end{center}
\end{figure}

It is, in principle, possible to extract from the measured geo-neutrino signal the Earth's radiogenic heat power.
This procedure is however not straightforward: the geoneutrino flux depends not only on the total mass of HPE in the Earth, but also on their distributions, which is model dependent. The HPE abundances and so the radiogenic heat in the crust are rather well known, as discussed in Secs.~\ref{Sec:Earth} and~\ref{Sec:signal}. As the main unknown remains the radiogenic power of the Earth's mantle. Figure~\ref{Fig:KBTW} summarizes the analysis we have performed in order to extract the mantle radiogenic heat from the measured geo-neutrino signals.

The geo-neutrino luminosity $\Delta L$ ($\bar{\nu}_e$ emitted per unit time from a volume unit, so called {\it voxel}) is related~\cite{fiorentini2007} to the U and Th masses $\Delta m$ contained in the respective volume:
\begin{equation}
\Delta L =7.46 \cdot \Delta m(^{238}\rm {U}) +1.62 \cdot \Delta m(^{232}{\rm Th})
\label{Eq:Lm}
\end{equation}
where the masses are expressed in units of 10$^{17}$~kg, and the luminosity in units of 10$^{24}$\,s$^{-1}$.

The measured geo-neutrino signal at a given site can be deduced by summing up the U and Th contributions from individual voxels over the whole Earth, and by weighting them by the inverse squared-distance (geometrical flux reduction) and by the oscillation survival probability. We have performed such an integration for the mantle contribution to the geo-neutrino signal. We have varied the U and Th abundances (with a fixed chondritic mass ratio Th/U = 3.9) in each voxel. The homogeneous and sunken layer models of the HPE distributions in the mantle (Sec.~\ref{Sec:signal}) were taken into account separately. For each iteration of different U and Th abundances and distributions, the total mantle geo-neutrino signal (taking into account Eq.~\ref{Eq:Lm}) and the U +Th radiogenic heat power from the mantle (considering Eq.~4 from~\cite{fiorentini2007}) can be calculated. The result is shown in Figure~\ref{Fig:KBTW} showing the U + Th mantle radiogenic heat power as a function of the measured mantle geo-neutrino signal. The solid lines represent the sunken-layer model, while the dotted lines the homogeneous mantle. The individual U and Th contributions, as well as their sums are shown. The measured mantle signal $S_{Mantle}^{SYM} = (7.7 \pm 6.2)~{\rm TNU}$ from the combined Borexino and KamLAND analysis quoted in Eq.~\ref{Eq:homo} is demonstrated on this plot by the vertical solid (orange) line indicating the central value of 7.7\,TNU while the filled (light brown) triangular area corresponds to $\pm 6.2$\,TNU band of 1$\sigma$ error. The central value of $S_{Mantle}^{SYM}$ = 7.7\,TNU corresponds to the mantle radiogenic heat from U and Th of 7.5 - 10.5\,TW (orange double arrow on $y$-axis), for sunken-layer and homogeneous HPE extreme distributions, respectively. If the error of the measured mantle geo-neutrino signal is considered ($\pm 6.2$\,TNU),  the corresponding interval of possible mantle radiogenic heat is from 2 to 19.5\,TW, indicated by the black arrow on $y$-axis.

\section{Conclusions and future perspectives}
\label{Sec:future}

The two independent geo-neutrino measurements from the Borexino and KamLAND  experiments have opened the door to a new inter-disciplinary field, the Neutrino Geoscience. They have shown that we have a new tool for improving our knowledge on the HPE abundances and distributions.
The first attempts of combined analysis has appeared~\cite{lisi,fiorentini2012,gando2011,BXgeo2}, showing the importance of multi-site measurements.  The first indication of a geo-neutrino signal from the mantle has emerged.  The present data seem to disfavor the geo-dynamical BSE models, in agreement with the recent understanding of the important role of water in the heat transportation engine.

These results together with the first attempts to directly measure the Th/U ratio are the first examples of geologically relevant outcomes.
But in order to find definitive answers to the questions correlated to the radiogenic heat and HPE abundances, more data are needed. 
Both Borexino and KamLAND experiments are going on to take data and a new generation of experiments using liquid scintillators is foreseen. 
One experimental project, SNO+ in Canada, is in an advanced construction phase, and a new ambitious project, Daya-Bay~2 in China, mostly aimed to study the neutrino mass hierarchy, has been approved. Other interesting proposals have been presented, LENA at Pyh\"asalmi (Finland) or  Fr\'ejus (France) and Hanohano in Hawaii.

The SNO+ experiment in the Sudbury mine in Canada~\cite{sno1,sno2}, at a depth of 6080 m.w.e.,  is expected to start the data-taking in 2014 - 2015. The core of the detector is made of $\sim$780 ton of LAB (linear alkylbenzene) with the addition of PPO as fluor. A rate of $\sim$20 geo-neutrinos/year is expected and the ratio of geo-neutrino to reactor $\bar{\nu}_e$ events should be around $\sim$1.2.
The site is located on an old continental crust and it contains significant quantities of felsic rocks, which are enriched in U and Th. Moreover, the crust is particularly thick (ranging between 44.2 km and 41.4 km),  approximately 40\% thicker than the crust surrounding the Gran Sasso and Kamioka sites. For these reasons, a strong LOC signal is expected, around 19 TNU. A very detailed study of the local geology is mandatory to allow the measurement of the mantle signal.

The main goal of the Daya Bay 2 experiment in China~\cite{dayabay2} is to determine the neutrino mass hierarchy. Thanks to a very large mass of 20\,kton it would detect up to 400 geo-neutrinos per year. A few percent precision of the total geo-neutrino flux measurement could be theoretically reached within the first couple of years and the individual U and Th contributions could be determined as well. Unfortunately, the detector site is placed by purpose very close to the nuclear power plant. Thus, under the normal operating conditions, the reactor $\bar{\nu}_e$ flux is huge ($\sim$40 detected events/day). Data interesting for the geo-neutrino studies could be probably taken only in correspondence with reactor maintenance or shutdowns.

 LENA is a proposal for a huge, 50\,kton liquid scintillator detector aiming to the geo-neutrino measurement as one of the main scientific goals~\cite{lena}. Two experimental sites have been proposed: Fr\'ejus in France or Pyh\"asalmi in Finland. From the point of view of the geo-neutrino study, the site in Finland would be strongly preferable, since Fr\'ejus is very close to the French nuclear power plants. LENA would detect about 1000 geo-neutrino events per year: a few percent precision on the geo-neutrino flux could be reached within the first few years, an order of magnitude improvement with
respect to the current experimental results. Thanks to the large mass, LENA would be able to measure the Th/U ratio, after 3 years with 10-11\% precision in Pyh\"asalmi and 20\% precision in Fr\'ejus.

Another very interesting project is Hanohano~\cite{hanohano} in Hawaii, placed on a thin, HPE depleted oceanic crust. The mantle contribution to the total geo-neutrino flux should be dominant, $\sim$75\%. A tank of 26\,m in diameter and 45\,m tall, housing a 10\,kton liquid scintillator detector, would be placed vertically on a 112\,m long barge and deployed in the deep ocean at 3 to 5\,km depth. The possibility to build a recoverable and portable detector is part of the project. A very high geo-neutrino event rate up to about $\sim$100 per year would be observed with a geo-neutrino to reactor-$\bar{\nu}_e$ event rate ratio larger than 10.
  
In conclusion, the new inter-disciplinary field has formed. The awareness of the potential to study our planet with geo-neutrinos is increasing within both geological and physical scientific communities  This is may be the key step in order to promote the new discoveries about the Earth and the new projects measuring geo-neutrinos.

\section*{Conflict of Interest}

The authors declare that there is no conflict of interests regarding the publication of this article.

\end{document}